\documentclass[aps,floats,amssymb,amsmath,prb,10pt,twocolumn,superscriptaddress,longbibliography]{revtex4-2}

\usepackage[utf8]{inputenc}
\usepackage{graphicx} %
\usepackage{amsmath}
\usepackage{color}
\usepackage{units}
\usepackage{braket}
\usepackage{newunicodechar}
\usepackage[colorlinks=true,allcolors=blue]{hyperref}

\newcommand{\sigmadc}{\sigma_\mathrm{dc}}

\begin{document}

\title{Quantum and classical transport in the underscreened Kondo lattice}

\author{Leo Pintar}
\affiliation{École polytechnique, Institut Polytechnique de Paris, Route de Saclay, 91128 Palaiseau Cedex, France}

\author{Don Rolih}
\affiliation{Jo\v{z}ef Stefan Institute, Jamova 39, SI-1000 Ljubljana, Slovenia}
\affiliation{Faculty  of Mathematics and Physics, University of Ljubljana, Jadranska 19, SI-1000 Ljubljana, Slovenia}

\author{Rok Žitko}
\email{rok.zitko@ijs.si}
\affiliation{Jo\v{z}ef Stefan Institute, Jamova 39, SI-1000 Ljubljana, Slovenia}
\affiliation{Faculty  of Mathematics and Physics, University of Ljubljana, Jadranska 19, SI-1000 Ljubljana, Slovenia}

\date{November 23, 2026}

\begin{abstract}
We investigate quantum effects in high-spin Kondo-lattice transport as
resistivity approaches the classical spin-disorder limit.
Using dynamical mean-field theory and the numerical renormalization group,
we calculate resistivity, the Seebeck coefficient, and optical conductivity for the
paramagnetic single-channel Kondo lattice away from half filling.
At intermediate temperatures, the resistivity for large local spin $S$ approaches
the double-exchange plateau. Thermopower
retains a positive
quantum excess even where resistivity is nearly classical. The excess arises from
the transport kernel's particle-hole asymmetry. Even the full Kelvin approximation,
often accurate for incoherent correlated metals, captures only part of it.
Incomplete Kondo screening leaves residual
moments and leads to singular Fermi-liquid behavior at low temperatures. For $S=1$,
the apparent power-law resistivity is consistent with a $1/\ln^2(T_\rho/T)$ 
asymptote. Above the Kondo scale $T_K$, weak-coupling resistivity follows
$1/\ln^2(T/T_K)$. Above the band and exchange scales, the charge susceptibility
decreases as $1/T$, while the effective diffusion constant becomes nearly
temperature independent, resulting in linear resistivity.
A many-body coherent-potential approximation accounts for the spin dependence of
the linear coefficient without adjustable parameters.
\end{abstract}

\maketitle

\section{Introduction}
\label{sec1}

In the Kondo lattice model (KLM), local moments of spin $S$ couple to itinerant
electrons through antiferromagnetic exchange $J>0$. The spin $S$ tunes the model
between two limits. At finite $S$, spin-flip exchange makes the moments dynamical
and leads to Kondo screening at low temperatures. As $S\to\infty$ at fixed $JS$,
the exchange-field scale stays fixed while quantum spin corrections, of relative
order $1/S$, vanish. The electrons, which remain quantum mechanical, then scatter
elastically from static, randomly oriented classical spins \cite{Furukawa1994}.
We investigate how the accuracy of this classical-spin description depends on
temperature and on the transport observable.

For $S=1/2$, the KLM describes the low-energy physics of the periodic Anderson
model for heavy-fermion compounds in the Kondo-lattice regime, where the $f$
orbitals are nearly singly occupied. Complete screening produces a coherent heavy
Fermi liquid in the paramagnetic metal. Its resistivity evolves from a low-temperature
$T^2$ law through a coherence maximum to incoherent Kondo scattering
\cite{Lavagna1982,Logan2005,Grenzebach2006}. Thermopower has a maximum sensitive to
band filling and particle-hole asymmetry \cite{Lavagna1982thermopower,Grenzebach2006},
while in the optical conductivity coherence produces a narrow Drude peak and an
interband peak from transitions across the hybridization gap
\cite{TahvildarZadeh1999,Logan2005,Grenzebach2006}. Dynamical mean-field theory
(DMFT) finds that low-energy transport at fixed filling is governed by a single
coherence scale \cite{Logan2005,Grenzebach2006}.

A single conduction band cannot fully screen moments with $S>1/2$. On an occupied site, a strong antiferromagnetic Kondo coupling forms a composite multiplet with total spin $S-1/2>0$, rather than a Kondo singlet. The low-energy manifold thus contains partially screened $S-1/2$ moments on occupied sites and unscreened spin-$S$ moments on empty sites. Electron hopping transfers the partially screened configuration between neighboring sites already at first order in the hopping amplitude $t$, making these residual-spin degrees of freedom mobile \cite{Masui2022}.

Underscreening changes the impurity fixed point. The residual moment remains ferromagnetically coupled to the conduction electrons through a marginally irrelevant interaction that vanishes only logarithmically at low energies. The resulting state is a singular Fermi liquid, characterized by elastic scattering at the Fermi level but nonanalytic low-energy corrections to conventional Fermi-liquid behavior \cite{Coleman2003,Mehta2005}. Logarithmic conductance associated with underscreening has been observed in molecular transistors \cite{Roch2009}; there it arises from logarithmic corrections to the unitary conductance of an underscreened quantum dot \cite{Posazhennikova2005}. On the lattice, the residual moments provide additional low-energy degrees of freedom that can develop collective correlations and magnetic order. A large-$N$ treatment of the underscreened Kondo lattice found singular dynamics and pseudogap formation in place of the conventional coherent heavy Fermi liquid \cite{Florens2004}, while studies of the $S=1$ model have focused primarily on its magnetic properties and possible relevance to uranium heavy-fermion compounds \cite{Perkins2007}. Previous DMFT studies of the $S=1$ Kondo lattice examined ferromagnetic ordering and Lifshitz transitions at low conduction-band filling \cite{Golez2013}, and the robustness of spectral features in the antiferromagnetic phase at half filling \cite{Osolin2015}. Recent gauge mean-field work maps out the ground-state phase diagram of the $S=1$ Kondo lattice, including a phase in which magnetic order coexists with Kondo hybridization \cite{Scott2026}. Despite these developments, the finite-temperature transport properties of the paramagnetic single-channel underscreened Kondo lattice remain little explored.

Related slow spin dynamics occur in the impurity models underlying Hund metals.
In the three-orbital Anderson model with Kanamori interaction and two electrons,
the effective Kondo Hamiltonian contains a spin exchange that can be much weaker
than the orbital and mixed spin-orbital couplings, or even
ferromagnetic. This
separates the screening scales: the orbital moment is quenched
while the spin remains active to much lower temperatures \cite{Horvat2016}.
That model ultimately reaches a regular
Fermi-liquid fixed point. A closer connection arises in orbital-selective Mott
phases, where localization of one orbital leaves a moment coupled by
ferromagnetic Hund exchange to itinerant electrons. Their effective
ferromagnetic Kondo-lattice description connects multiorbital Hubbard models
to double exchange \cite{Biermann2005}, the strong-coupling limit in which
the itinerant-electron spin follows the local moments, so that hopping
between sites depends on the relative orientation of their moments
\cite{AndersonHasegawa1955}. DMFT studies find underscreened
impurity physics and logarithmic self-energies in the remaining metallic
orbitals \cite{Greger2013,Kugler2019}. Interorbital hopping can restore
screening at a very low coherence scale \cite{KuglerKotliar2022}.

In the infinite-coordination limit, DMFT maps the Kondo lattice onto a self-consistent $s-d$ impurity problem \cite{Georges1992,Matsumoto1995,Han2021}, treating local Kondo dynamics nonperturbatively with lattice feedback. In this work, we apply DMFT with a numerical renormalization group (NRG) impurity solver to the paramagnetic single-channel high-spin Kondo lattice at generic filling, calculating the resistivity, Seebeck coefficient, and optical conductivity over a broad range of temperatures.
Our results show how quantum spin effects in transport weaken with increasing $S$,
giving way to classical spin-disorder scattering.
Thermopower retains a quantum excess beyond both classical and Kelvin
descriptions, and at temperatures above the band and exchange scales, a
many-body coherent-potential approximation explains the linear resistivity.

\section{Summary of results}

We summarize the transport regimes in order of increasing temperature.

At low temperatures, the underscreened models exhibit apparent power-law
resistivities over finite temperature windows. For $S=1$, we relate the drifting exponent to
marginally irrelevant ferromagnetic scattering and derive an inverse-log-squared
asymptote with a two-loop logarithmic correction. Optical conductivity also
distinguishes full from partial screening: the fully screened lattice displays the
non-Drude ``foot'' of a local Fermi liquid \cite{Berthod2013}, absent in the
underscreened models. Above the Kondo scale $T_K$,
weak-coupling resistivity follows $1/\ln^2(T/T_K)$. These logarithmic
quantum corrections weaken with increasing $S$, leaving an increasingly broad
window of nearly classical spin-disorder transport.

The approach to classical behavior depends on the observable. At intermediate
temperatures, the large-spin resistivity approaches the classical double-exchange
value \cite{Furukawa1994} once $JS$ becomes comparable to the half-bandwidth $D$,
even at modest $J$, while thermopower retains a positive quantum excess. The full Kelvin approximation captures
only part of this excess and, for $S=1$, only about half of the thermopower at
every filling studied. This is a notable exception: the Kelvin approximation
describes thermopower well for incoherent transport in the Hubbard model and in
the Hund metal Sr$_2$RuO$_4$ \cite{Arsenault2013Thermopower,Mravlje2016}, and it
is exact for Sachdev--Ye--Kitaev models \cite{Davison2017}. Similar deviations due to
asymmetric scattering are found in skewed non-Fermi liquids \cite{GeorgesMravlje2021,Gourgout2022}.

At temperatures exceeding the band and exchange scales, the resistivity becomes linear in $T$.
The charge susceptibility decreases as $\chi_c\propto1/T$, while the effective diffusion constant
extracted from conductivity and susceptibility becomes nearly temperature independent.
This connects the Kondo lattice to the thermodynamic mechanism identified in high-temperature
Hubbard-model transport \cite{Perepelitsky2016,Kokalj2017}. The spin multiplicity cancels from the charge
statistics in that limit: $T\chi_c$ and thermopower become spin independent, while the transport
kernel retains the spin dependence of the exchange multiplets.
We account for the spin dependence of the linear coefficient using the
many-body coherent-potential approximation (CPA) of Ref.~\cite{Edwards1999}, specialized to
the infinite-temperature atomic weights and antiferromagnetic exchange. 
The model separates the thermodynamic origin of $T$-linear resistivity
from the microscopic dynamics that determine its coefficient.

\section{Model}

We consider the single-channel Kondo lattice model with isotropic antiferromagnetic exchange,
\begin{equation}
    H_K = \sum_{k\sigma} \epsilon_k c^\dag_{k\sigma} c_{k\sigma} + J \sum_i \mathbf{s}_i \cdot \mathbf{S}_i.
\end{equation}
Here $c_{k\sigma}$ annihilates a conduction electron with momentum $k$ and spin $\sigma=\uparrow,\downarrow$, and $\epsilon_k$ is its bare dispersion measured from the band center. The conduction-electron spin density is $\mathbf{s}_i=(1/2)\sum_{\alpha\beta}c^\dag_{i\alpha}\boldsymbol{\sigma}_{\alpha\beta}c_{i\beta}$, where $\alpha,\beta\in\{\uparrow,\downarrow\}$ and $\boldsymbol{\sigma}$ is the vector of Pauli matrices. The localized moments are quantum spins with $\mathbf{S}_i^2=S(S+1)$, and $J>0$ is the exchange coupling.

The noninteracting band is specified by the semicircular density of states of the infinite-coordination Bethe lattice,
\begin{equation}
\nu_0(\epsilon)=\frac{2}{\pi D^2}\sqrt{D^2-\epsilon^2},
\qquad |\epsilon|\le D.
\end{equation}
The DOS is normalized to one per conduction-electron spin. The half-bandwidth is $D$. We set $D=1$ and use $k_B=\hbar=1$, expressing temperatures in energy units.

The filling $n=(1/N)\sum_{i\sigma}\langle c^\dag_{i\sigma}c_{i\sigma}\rangle$ is the conduction-electron number per site, including both spins, with $N$ the number of sites and $0\le n\le2$. It is controlled by the chemical potential $\mu$, which enters the grand-canonical Hamiltonian as $H_K-\mu\sum_{i\sigma}c^\dag_{i\sigma}c_{i\sigma}$. We focus on the paramagnetic phase below half filling, $0<n<1$, with $n=0.8$ unless stated otherwise. For the symmetric band, particle--hole symmetry maps $(\mu,n)$ to $(-\mu,2-n)$. The resistivity and optical conductivity are unchanged under this transformation, while the Seebeck coefficient $\mathcal S$ changes sign, $\mathcal S(T,2-n)=-\mathcal S(T,n)$, and vanishes at half filling.

\section{Methods}

\subsection{DMFT}
\label{sec:dmft}

We solve the KLM within single-site DMFT, restricting the solution to the homogeneous paramagnetic
phase. In this approximation, which becomes exact in the infinite-coordination limit, the
conduction-electron self-energy is local, $\Sigma_{\mathbf{k}}(z)=\Sigma(z)$
\cite{Georges1992,Matsumoto1995}. The lattice is mapped onto an impurity consisting of a local
conduction orbital exchange-coupled to spin $S$ and hybridized with a noninteracting bath. This
permits direct evaluation of the conduction-electron self-energy \cite{Osolin2015}. We use the implementation from Ref.~\cite{DMFTNRGKLM}.

The local lattice Green's function is
\begin{equation}
G(z)=\int_{-D}^{D}d\epsilon\,
\frac{\nu_0(\epsilon)}
{z+\mu-\epsilon-\Sigma(z)}.
\label{eq:dmft_lattice}
\end{equation}
Here $z$ is a complex frequency, with $z=\omega+i0^+$ for the retarded functions and $\omega$ measured relative to $\mu$. The bath is described by the hybridization function $\Delta(z)$, and self-consistency requires $G_{\rm imp}(z)=G(z)$. For the semicircular DOS, the corresponding relations are
\begin{equation}
\begin{aligned}
  G_{\rm imp}^{-1}(z)&=z+\mu-\Delta(z)-\Sigma(z),\\
  \Delta(z)&=\frac{D^2}{4}G(z).
\end{aligned}
\label{eq:dmft_bethe}
\end{equation}
The local spectrum $A_{\rm loc}=-\operatorname{Im}G^R/\pi$ and the band DOS $\nu_0$ are defined per conduction-electron spin. The filling is therefore $n=2\int d\omega\,f(\omega)A_{\rm loc}(\omega)$, where $f(\omega)=[e^{\omega/T}+1]^{-1}$.

The self-consistency equations are solved directly on the real-frequency axis. We stabilize the iteration by linearly mixing $\Delta$, with a standard weight $\alpha_{\rm mix}=0.5$ for the new bath. The real-frequency mesh is refined to resolve structure in the hybridization spectrum $\Gamma_\Delta(\omega)=-\operatorname{Im}\Delta^R(\omega)$. To maintain fixed filling, we adjust $\mu$ during the iteration. In the standard update, the filling at a trial $\mu+\delta\mu$ is estimated by rigidly translating the current spectrum, $A_{\rm loc}(\omega)\to A_{\rm loc}(\omega+\delta\mu)$. The resulting chemical-potential step is underrelaxed and bounded. Computing trial fillings from the lattice Green function at fixed self-energy gives consistent results, but rigid shifting converges slightly faster and more reliably in our calculations.

Convergence is monitored using the integrated absolute difference between successive lattice spectra,
\begin{equation}
  r_k=\int d\omega\,
  \left|A_{\rm loc}^{(k)}(\omega)-A_{\rm loc}^{(k-1)}(\omega)\right|,
  \label{eq:dmft_residual}
\end{equation}
where $k$ labels the DMFT iteration and the spectra are compared on a common mesh. We require $r_k<\varepsilon_{\rm DMFT}$ over a prescribed number of consecutive iterations. Our standard threshold is $\varepsilon_{\rm DMFT}=10^{-5}$; the threshold, iteration window, and mixing are adjusted for the different parameter regimes. 

\subsection{NRG}
\label{sec:nrg}

The impurity problem is solved by the numerical renormalization group
\cite{Wilson1975,KrishnaMurthy1980,Bulla2008}, using the \textsc{NRG Ljubljana} code
\cite{NRGLjubljana}. NRG provides logarithmic resolution of small Kondo scales
and computes real-frequency spectra without analytic continuation. We exploit
$\mathrm{U}(1)$ charge and $\mathrm{SU}(2)$ spin symmetries and use the improved discretization of Ref.~\cite{Zitko2009}. Our standard logarithmic discretization parameter is $\Lambda=2$, with $N_z=4$ interleaved grids. At iteration $N$, we keep states with excitation energies below $10E_N$, where $E_N$ is the characteristic NRG energy scale. We impose a nominal maximum of 10000 symmetry multiplets and a minimum of 500 when available.

Finite-temperature spectral functions and expectation values are evaluated using full-density-matrix NRG \cite{Weichselbaum2007}. Our standard reconstruction uses a modified log-Gaussian kernel with broadening parameter $\alpha=0.3$, followed by convolution with a Fermi-derivative kernel of temperature scale $0.2T$. The broadened spectra are evaluated on the adaptive real-frequency mesh. We reconstruct the real parts of the impurity Green's function and auxiliary correlators by Kramers--Kronig transformation and evaluate the improved $\Sigma$ estimator of Ref.~\cite{Kugler2022}.

The truncation, $z$ averaging, broadening, frequency mesh, and DMFT stopping parameters are adapted to the spin and temperature regime. The low-temperature spectra in Fig.~\ref{fig3} use $N_z=16$ and $\alpha=0.2$. For the high-temperature resistivity and spectra in Fig.~\ref{fig:high_temperature}(a,b), we use $N_z=16$, $\alpha=0.03$, and $\varepsilon_{\rm DMFT}=10^{-4}$.
The significantly lower value of $\alpha$ for $T \gg D$ is critically important to correctly
recover the asymptotic linear-in-$T$ resistivity. Overbroadening manifests as an underestimation
of the computed resistivity at high temperatures (downward curvature).

For $S=1/2$, resolving the vanishing $\mathrm{Im}\Sigma(\omega,T)$ as $\omega,T\to0$
is numerically challenging in DMFT(NRG). The improved self-energy
estimator \cite{Kugler2022} substantially reduces causality violations,
but residual numerical artifacts produce spurious saturation of $\rho$
at very low temperatures.
Reducing the discretization parameter $\Lambda$ systematically extends the
temperature range of clear $T^2$ scaling, even at fixed broadening $\alpha$;
reducing $\alpha$ improves the results further. Approaching the
continuum limit is very costly.

\subsection{Transport calculations}
\label{sec:transport}

We calculate the transport coefficients from the current bubble \cite{Zitko2015,Ding2017}. For a single-band, inversion-symmetric lattice, vertex corrections to this response vanish in the infinite-dimensional limit. The spectral function entering the bubble is
\begin{equation}
  A(\omega,\epsilon)=-\frac{1}{\pi}\operatorname{Im}
  \frac{1}{\omega+\mu-\epsilon-\Sigma^R(\omega)}.
  \label{eq:transport_spectrum}
\end{equation}
Here $\epsilon$ is the bare band energy measured from the band center. We write the local spectrum as $A(\omega)\equiv A_{\rm loc}(\omega)=\int_{-D}^{D}d\epsilon\,\nu_0(\epsilon)A(\omega,\epsilon)$.
For the Bethe lattice, the transport function consistent with the optical sum rule is \cite{Arsenault2013Transport,Zitko2015,Ding2017}
\begin{equation}
  \Phi(\epsilon)=(D^2-\epsilon^2)\nu_0(\epsilon)
  =\frac{2D}{\pi}\left[1-(\epsilon/D)^2\right]^{3/2}
  \label{eq:transport_function}
\end{equation}
for $|\epsilon|\le D$. Geometrical and electromagnetic prefactors are absorbed into the conductivity unit $\sigma_0$; we quote resistivity in units of $\rho_0=1/\sigma_0$. In terms of the physical transport function $\Phi_{xx}(\epsilon)=V^{-1}\sum_k(\partial\epsilon_k/\partial k_x)^2\delta(\epsilon-\epsilon_k)$, with volume $V$ and no spin sum, the convention is $\sigma_0=e_0^2\Phi_{xx}(0)/(\hbar D)$ and $\Phi_{xx}(\epsilon)/\Phi_{xx}(0)=\Phi(\epsilon)/\Phi(0)$. The regular real part of the optical conductivity is then
\begin{equation}
\begin{aligned}
  \frac{\sigma_1(\Omega,T)}{\sigma_0}
  &={}\pi^2\int_{-\infty}^{\infty}d\omega\,
  \frac{f(\omega)-f(\omega+\Omega)}{\Omega}\\
  &\quad\times\int_{-D}^{D}d\epsilon\,\Phi(\epsilon)
  A(\omega,\epsilon)A(\omega+\Omega,\epsilon).
\end{aligned}
\label{eq:optical_bubble}
\end{equation}
We denote the photon energy by $\Omega$ and write $f(\omega)=[e^{\omega/T}+1]^{-1}$, with $T$ in energy units.
Both conduction-electron spin projections are included. This expression coincides with the normalization used by the \textsc{Bubble} code, which we use to evaluate the transport integrals from the real-frequency self-energy.\footnote{\protect\url{https://github.com/rokzitko/bubble}.}

In the dc limit, $[f(\omega)-f(\omega+\Omega)]/\Omega\to-f'(\omega)$. Defining the transport kernel and its moments by
\begin{equation}
\begin{aligned}
  \mathcal K(\omega,T)&=\int_{-D}^{D}d\epsilon\,
  \Phi(\epsilon)A(\omega,\epsilon,T)^2,\\
  L_j(T)&=\int_{-\infty}^{\infty}d\omega\,
  [-f'(\omega)]\,\omega^j\mathcal K(\omega,T),
  \quad j=0,1,
\end{aligned}
\label{eq:transport_moments}
\end{equation}
we obtain 
\begin{equation}
  \frac{\rho_\mathrm{dc}}{\rho_0}=\frac{1}{\pi^2L_0},
  \qquad
  \frac{\mathcal S}{k_B/e_0}=-\frac{L_1}{TL_0}.
  \label{eq:dc_transport}
\end{equation}
Here $\mathcal S$ is the Seebeck coefficient and $e_0>0$ is the magnitude of the electron charge. We write $\rho\equiv\rho_\mathrm{dc}=1/\sigmadc$, with $\sigmadc=\lim_{\Omega\to0}\sigma_1(\Omega)$.

To characterize the low-frequency response, we compare it with the Drude form
\begin{equation}
  \sigma_{D,1}(\Omega)=\frac{\sigma_{D,\mathrm{dc}}}
  {1+(\Omega/\gamma_D)^2}.
\label{eq:drude_reference}
\end{equation}
Here $\gamma_D=\hbar/\tau_D$ is the energy half-width at half-maximum, with $\tau_D$ the Drude
relaxation time.

The fully screened metallic $S=1/2$ case provides the regular Fermi-liquid reference. In its coherent regime, the quasiparticle contribution is
\begin{equation}
  G^R(k,\omega)\simeq\frac{Z}{\omega-E_k^*+i\gamma_\mathrm{qp}},
  \label{eq:qp_reference}
\end{equation}
where $Z=[1-\partial_\omega\operatorname{Re}\Sigma^R(0)]^{-1}$, $E_k^*=Z[\epsilon_k+\operatorname{Re}\Sigma^R(0)-\mu]$, and $\gamma_\mathrm{qp}=-Z\operatorname{Im}\Sigma^R(0)$ is the single-particle spectral half-width. Convolving two Lorentzians with approximately constant width and a smooth transport function gives $\gamma_D\simeq2\gamma_\mathrm{qp}=-2Z\operatorname{Im}\Sigma^R(0)$.

\section{Overview of spectral and transport properties}
\label{sec:overview}

\begin{figure}[t]
    \centering
    \includegraphics[width=\columnwidth]
    {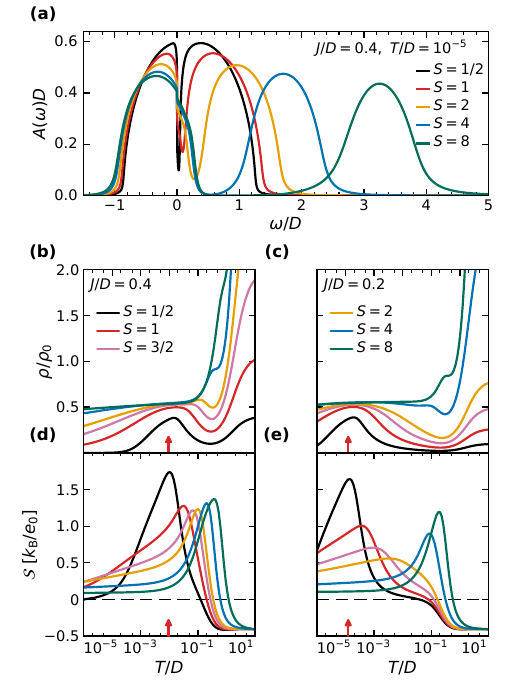}
    \caption{
    Overview of spectral and transport properties at filling $n=0.8$.
    (a) Conduction-band spectral functions $A(\omega)D$ at $J/D=0.4$
    and $T/D=10^{-5}$.
    (b,c) Normalized resistivity $\rho/\rho_0$ and
    (d,e) Seebeck coefficient in units of $k_{\mathrm B}/e_0$ versus temperature,
    for $J/D=0.4$ (left) and $J/D=0.2$ (right).
    Red arrows on the temperature axes of (b--e) mark the
    weak-coupling Kondo-scale estimates (see text).
    Thin dashed lines in (d,e) mark zero thermopower.
    }
    \label{fig1}
\end{figure}

Figure~\ref{fig1} summarizes spectral and transport properties at $n=0.8$. The local spectra in panel (a), calculated at $J/D=0.4$ and $T/D=10^{-5}$, exhibit a dip
at positive frequency, with the Fermi level on its lower-energy flank. For the smaller spins, this
depletion is associated with Kondo screening and the resulting hybridization structure. As $S$
increases, the upper spectral peak moves to higher energy and the dip develops into an extended,
gap-like interval above the Fermi level, while $A(0)$ remains finite. The increasingly separated
bands can be understood from the local one-electron multiplets: $S_T=S-1/2$ with energy
$E_-=-J(S+1)/2$, and $S_T=S+1/2$ with energy $E_+=JS/2$ \cite{Masui2022}. Their splitting is $\Delta_{\rm ex}=J(S+1/2)$. For $S=8$, the spectral maxima are indeed separated by approximately $3.6D$, close to the atomic splitting $3.4D$.
For classical spins, the exchange field splits the band into two subbands at
$\mp JS/2$, separated by $JS$ and with equal weights. Quantum spins shift the
lower subband down by $J/2$ and give the two multiplets unequal degeneracies,
$2S$ and $2S+2$. Analogous quantum corrections occur for
ferromagnetic coupling \cite{Meyer2001}, with the order of the multiplets reversed.
As $JS$ increases, the hybridization picture evolves continuously into a
description in terms of local exchange multiplets.

Panel (b) shows the temperature dependence of resistivity at $J/D=0.4$. For $S=1/2$, the low-temperature suppression reflects Fermi-liquid coherence, with asymptotic resistivity $\rho\propto T^2$. On warming, the resistivity passes through a maximum near $T/D=0.014$ as lattice coherence is lost, decreases through the incoherent exchange-scattering regime, and reaches a minimum near $T/D=0.27$ before rising again. Increasing $S$ generally raises the low-temperature resistivity and progressively weakens this maximum--minimum structure. The maximum also shifts upward in temperature, reaching $T/D\simeq0.115$ for $S=2$. For $S=4$ and $8$, broad, nearly coincident plateaus with $\rho/\rho_0\simeq0.5$ replace the resolved maxima. This resistivity remains large down to the lowest temperatures shown.

Reducing the coupling to $J/D=0.2$, panel (c), moves the smaller-spin resistivity maxima to much lower temperatures while leaving their heights nearly unchanged. For $S\le3/2$, the maxima lie near $T/D=1.9\times10^{-4}$, and the subsequent decrease extends over a much broader temperature interval. 

The coupling dependence of the resolved resistivity maxima is consistent with the single-impurity
Kondo scale. Let $\epsilon_F$ be the noninteracting Fermi energy at filling $n$, with $\nu_F=\nu_0(\epsilon_F)$. At $n=0.8$, the bare-band Fermi-level DOS is $\nu_FD=0.629$, giving $g=\nu_FJ=0.251$ for $J/D=0.4$. We use the weak-coupling estimate $T_K^{\rm est}/D=\sqrt{g}\,e^{-1/g}$, with the order-one prefactor set to unity. This gives $T_K^{\rm est}/D\simeq0.0094$ for $J/D=0.4$ and $1.25\times10^{-4}$ for $J/D=0.2$. For the isotropic single-channel impurity, the leading weak-coupling form has no explicit $S$ dependence \cite{Zitko2008Anisotropy}, so $T_K^{\rm est}$ gives a common single-impurity screening scale against which the maxima for all spins can be compared.

The Seebeck coefficient, shown in panels (d,e), has a pronounced positive maximum for each spin.
The peak temperature increases with $S$, but its height first decreases and then grows
for the largest spins, most clearly at $J/D=0.2$, where it is smallest for $S=2$. For larger $S$, the thermopower maximum
does not track the nominal single-impurity Kondo scale. Its upward shift with $S$ is consistent
with the growing exchange splitting and associated redistribution of spectral weight.

Across the resistivity plateau, the thermopower is far from constant: for $S=8$
at $J/D=0.4$, $\rho$ changes by about 5\% between $T/D=0.003$ and $0.05$, while
$\mathcal S$ nearly quadruples. Since thermopower probes the particle-hole asymmetry of the
transport kernel $\mathcal K$ of Eq.~\eqref{eq:transport_moments}, its
maximum need not coincide with
the resistivity maximum. Both spectral-weight redistribution and interaction-induced
electron--hole lifetime asymmetry contribute \cite{Deng2013,Zitko2015,Zitko2018Asymmetry}.

Beyond the positive maximum, the Seebeck coefficient changes sign, with the zero crossing moving to higher temperature as $S$ increases. At the highest temperatures the curves approach a common negative value, the spin-independent limit discussed in Sec.~\ref{sec:charge_linear}.

\section{Low-temperature regime: logarithmic or power-law asymptotics?}
\label{sec:lowT}

\subsection{Logarithmic resistivity}
\label{sec:logarithmic_resistivity}

\begin{figure}[t]
    \centering
    \includegraphics[width=\columnwidth]{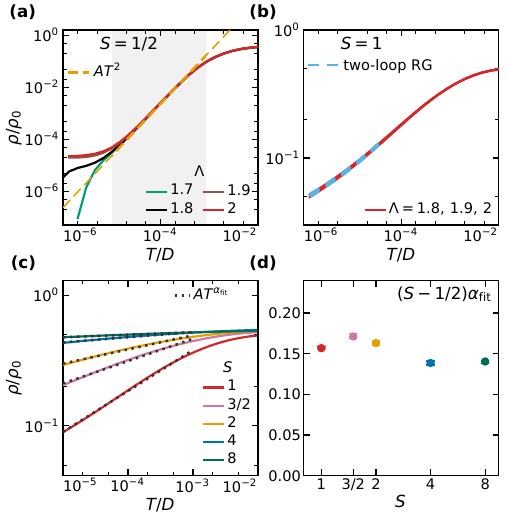}
    \caption{
Low-temperature resistivity $\rho/\rho_0$ at $J/D=0.4$ and filling $n=0.8$.
(a) $S=1/2$ DMFT results for several NRG discretizations
$\Lambda$ (solid curves).
The orange dashed $AT^2$ curve is fitted to the $\Lambda=1.8$ data.
The gray band marks the fitting interval.
(b) $S=1$ results (red).
Blue dashed lines are two-loop RG fits of Eq.~\eqref{eq:sfl_rho_fit}.
(c) Spin-dependent resistivity for $S\ge1$ (colored curves) and power-law fits
(gray dotted).
(d) Residual-spin-scaled exponents $(S-1/2)\alpha_{\rm fit}$ of those fits.
Fits in panels (b,c) are shown only over their fitting ranges.
    }
    \label{fig2}
\end{figure}

Figure~\ref{fig2}(a) illustrates the $S=1/2$ Fermi-liquid law $\rho\propto T^2$;
smaller $\Lambda$ extends the range of clean $T^2$ scaling (Sec.~\ref{sec:nrg}).
Numerical control is better for $S>1/2$: weak ferromagnetic scattering
from residual spins makes the damping $-\mathrm{Im}\Sigma$ vanish more slowly
as $\omega,T\to0$ than for $S=1/2$. For $S=1$, the resistivity curves
computed for $\Lambda=2$, 1.9, and 1.8 nearly overlap down to the lowest temperatures considered
[Fig.~\ref{fig2}(b)]. Although the resistivity resembles a power law,
its exponent drifts with temperature. This drift reflects
logarithmic behavior arising from the ferromagnetic coupling.

For $S=1$, at low energy the impurity has a residual spin $S_*=1/2$. Its
dimensionless exchange $g<0$ with the screening channel is marginally
irrelevant: reducing the excitation energy $E$ weakens the interaction,
but only logarithmically \cite{Koller2005,Mehta2005},
\begin{equation}
  g(E)\simeq-\frac{1}{\ln(T_{\rm s}/E)},\qquad E\ll T_K.
  \label{eq:sfl_running}
\end{equation}
The scale $T_{\rm s}$ characterizes the residual ferromagnetic
coupling and does not coincide with the Kondo screening scale $T_K$.
The electronic damping starts at order $g^2$.

The one-loop thermal function resolves how frequency and temperature cut
off the logarithm. Evaluating the electron--residual-spin loop underlying
the resummation of Ref.~\cite{CabreraCano2013} gives the thermal logarithm and the damping
\begin{equation}
\begin{aligned}
  \mathcal L(\omega,T)
  &=\ln\frac{T_{\rm s}}{2\pi T}
    -\operatorname{Re}\psi\!\left(\frac12-\frac{i\omega}{2\pi T}\right),\\
  \Gamma(\omega,T)&\equiv-\operatorname{Im}\Sigma^R(\omega,T)
    \simeq\frac{A_\Sigma}{\mathcal L(\omega,T)^2}.
\end{aligned}
\label{eq:sfl_thermal}
\end{equation}
Here $\psi$ is the digamma function, $\gamma_E$ is Euler's constant,
and $A_\Sigma>0$ has dimensions of
energy. For $T\ll|\omega|$, $\mathcal L\to\ln(T_{\rm s}/|\omega|)$, and at
$\omega=0$, $\mathcal L(0,T)=\ln[2e^{\gamma_E}T_{\rm s}/(\pi T)]$.
Equation~\eqref{eq:sfl_thermal} is a leading-logarithmic
approximation for $|\omega|,T\ll T_K$ and $\mathcal L\gg1$. Gauge mean-field
theory of the underscreened lattice likewise finds a logarithmically decaying
residual ferromagnetic interaction in the heavy-fermion metal \cite{Scott2026}.
Appendix~\ref{app:sfl_transport} presents a derivation of the thermal
function and its lattice version at finite filling.

DMFT self-consistency changes the parameters $A_\Sigma$ and $T_{\rm s}$ of
Eq.~\eqref{eq:sfl_thermal} but preserves its leading logarithms. 
In the low-temperature limit of the transport bubble,
$\sigmadc\propto\langle\Gamma^{-1}\rangle_T$, where
$\langle F\rangle_T=\int d\omega\,[-f'(\omega)]F(\omega,T)$.
Thus the zero-frequency damping alone is insufficient. Nevertheless, the
leading result is $\rho\propto1/\ln^2(T_{\rm s}/T)$, with no extra power
of $T$.

For quantitative fits we include the first logarithmically enhanced
correction, fixed by the two-loop flow \cite{AbrikosovMigdal1970,Fowler1971,Fritz2004}. The form used in
Fig.~\ref{fig2}(b) is
\begin{equation}
\begin{gathered}
  \frac{\rho_{\rm fit}(T)}{\rho_0}=\frac{A_\rho}{y(T)^2},\\[2pt]
  y(T)-\frac12\ln\!\bigl[y(T)+\tfrac12\bigr]
    =\ln\frac{T_\rho}{T}.
\end{gathered}
\label{eq:sfl_rho_fit}
\end{equation}
The amplitude $A_\rho>0$ and scale $T_\rho>0$ are fitted. The positive
solution $y$ is the inverse coupling inferred from resistivity and agrees
with $-1/g(T)$ to leading order (Appendix~\ref{app:two_loop}). Writing
$\ell_\rho=\ln(T_\rho/T)$, one has
$y=\ell_\rho+\tfrac12\ln\ell_\rho+\cdots$. The scale $T_\rho$ absorbs finite thermal
and observable-matching corrections. It therefore differs from $T_{\rm s}$ by a
nonuniversal factor of order one; at one loop,
$T_\rho=e^{\gamma_E+1}T_{\rm s}/(2\pi)\approx0.77\,T_{\rm s}$
[Eq.~\eqref{eq:sfl_one_loop_rho}]. Differentiating
Eq.~\eqref{eq:sfl_rho_fit} gives the local slope
$\alpha_{\rm eff}=d\ln\rho_{\rm fit}/d\ln T=2/y+1/y^2$.
Its slow drift toward zero as $T\to0$ accounts for the apparent power law;
ultimately, resistivity decreases more slowly than any positive power of $T$. For $S\geq3/2$ the same forms hold
with spin-dependent $A_\rho$ and $T_\rho$. At fixed residual exchange field,
the leading logarithmic flow rate scales as $1/S_*$, where $S_*=S-1/2$. This accounts for the
approximate constancy of $S_*\alpha_{\rm fit}=0.138$--$0.171$ in
Fig.~\ref{fig2}(d); see also Appendix~\ref{app:larger_spins}.
This regime is the high-temperature Kondo regime run in reverse, a
duality we return to in Sec.~\ref{sec:kondo_duality}.

\subsection{Spectral and optical signatures}
\label{sec:spectral_optical}

\begin{figure}
    \centering
    \includegraphics[width=\columnwidth]{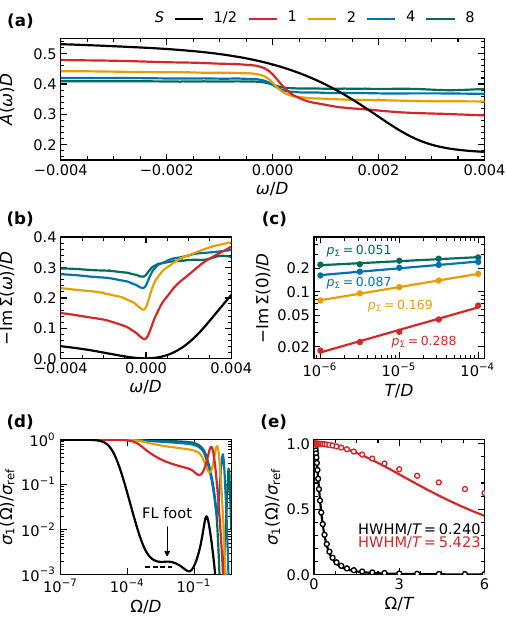}
    \caption{
    Low-temperature spectral and optical properties at $n=0.8$ and $J/D=0.4$.
    (a,b) Spectral function $A(\omega)D$ and $-\mathrm{Im}\Sigma(\omega)/D$
    close to the Fermi level.
    (c) $-\mathrm{Im}\Sigma(0)/D$ versus temperature.
    Symbols are numerical data; lines are log--log guides
    $-\mathrm{Im}\Sigma(0)\propto T^{p_\Sigma}$ over $10^{-6}\le T/D\le10^{-4}$.
    (d) Real part of the optical conductivity $\sigma_1(\Omega)$. The dashed line estimates the foot as
    $\sigmadc(\gamma_D/2\pi T)^2$, with the Drude half-width $\gamma_D$ from (e).
    (e) Drude-profile comparison for $S=1/2,1$ versus $\Omega/T$:
    symbols are numerical data and solid lines are Lorentzian fits.
    For $S=1$, the data exceed the Lorentzian above $\Omega\approx2T$.
    Panels (a,b,d,e) use $T/D=10^{-4}$, such that $T\ll T_K$.
    The optical curves in (d,e) show $\sigma_1(\Omega)/\sigma_{\rm ref}$,
    with $\sigma_{\rm ref}=\sigma_1(\Omega_{\rm ref})$ for each curve and
    $\Omega_{\rm ref}/D=10^{-6}$.
    The Drude fits use $10^{-2}\le\Omega/T\le1$.
    }
    \label{fig3}
\end{figure}

Figure~\ref{fig3}(a) shows the spectral anomalies accompanying the slow
decrease of $\rho(T)$ for large spins. At $T=10^{-4}D$, well below the nominal
$T_K$, the system lies in the regular ($S=1/2$) or singular ($S\geq1$) Fermi-liquid regime.
Despite the large $|\mathrm{Im}\Sigma(0)|$ for $S\geq1$
[Fig.~\ref{fig3}(b)], the spectral functions vary more sharply across
$\omega=0$ than in the regular Fermi liquid. Both the zero-frequency value
and the frequency dependence of $\mathrm{Im}\Sigma$ therefore matter.
For $S\geq1$, its cusp-like form is characteristic of a singular Fermi liquid.
A logarithmic cusp develops on cooling, but the approach to the
zero-temperature limit is logarithmically slow and resembles a power law
over a finite temperature window [Fig.~\ref{fig3}(c)].

For $S=1/2$, the optical conductivity [Fig.~\ref{fig3}(d)] reveals a
non-Drude signature of the heavy Fermi liquid. The
narrow Drude peak, of half-width $\gamma_D=0.24T$ [Fig.~\ref{fig3}(e)],
is followed for $2\pi T\lesssim\Omega\lesssim10^{-2}D$ by a nearly
frequency-independent shoulder, about three orders of magnitude below the Drude peak height.
This Fermi-liquid ``foot'' \cite{Berthod2013} follows from quadratic
frequency and temperature dependence of the scattering rate.
The weak maximum near $\Omega\approx6\times10^{-3}D$ marks the upper end
of the foot. At higher frequencies, intraband absorption is suppressed,
producing the dip at $\Omega\approx0.06D$. The satellite at
$\Omega\approx0.35D$ comes from direct interband transitions across the
hybridization gap. 

For $S\geq1$ the low-frequency response is not a conventional Drude
peak. The damping is large, $\Gamma(0)=0.07$--$0.28D$ for $S=1$--$8$,
far above the frequencies of interest, but the singular real part of
the self-energy, Eq.~\eqref{eq:sfl_real_part}, varies steeply within the
thermal window: for $S=1$,
$\operatorname{Re}\Sigma\simeq\mathrm{const}-C\tanh(\omega/2T)$ for
$|\omega|\lesssim2T$, with $\partial\operatorname{Re}\Sigma/\partial
\omega=-260$ at $\omega=0$. For $\Omega\lesssim2T$ this acts as a large
thermal mass renormalization. The finite-temperature slope factor
$Z_T=[1-\partial_\omega\operatorname{Re}\Sigma^R(0,T)]^{-1}\approx0.004$
rounds the response like a Drude peak of half-width $2Z_T\Gamma(0)\approx5T$
[cf.\ Eq.~\eqref{eq:qp_reference}], close to the
half-width of the Lorentzian fitted in Fig.~\ref{fig3}(e). At larger
$\Omega$, $\operatorname{Re}\Sigma$ becomes less steep and the
renormalization weakens, so $\sigma_1$ decays much more slowly than a
Lorentzian; for $S=1$ it reaches half height only at $\Omega\approx11T$.
Beyond this, $\sigma_1(\Omega)$ declines only through the growth of
$\Gamma(\omega)$ away from the Fermi level, Fig.~\ref{fig3}(b). At
$\Omega=10^{-3}D$ its local exponent
$d\ln\sigma_1/d\ln\Omega$ ranges from $-0.35$ for $S=1$ to $-0.018$ for $S=8$.
These magnitudes agree within about $6\%$ with the local
slopes of the dc resistivity at $T=10^{-4}D$, Fig.~\ref{fig2}(c), and
like them scale roughly as $1/S_*$ (Appendix~\ref{app:larger_spins}).
No foot appears for $S\geq1$.

\newcommand{\rhomax}{\rho_\mathrm{max}}

\section{Classical resistivity and quantum thermopower}
\label{sec:intermediate}

\subsection{Double-exchange plateau}
\label{sec:de_plateau}

The largest spins show a distinctive intermediate-temperature regime.
In Fig.~\ref{fig1}(b,c), the resistivity for $S=4$ and $8$ at
$J/D=0.4$, and for $S=8$ at $J/D=0.2$, stays close to $0.5\rho_0$ over
several decades of temperature instead of passing through the coherence
maximum of the small spins. The plateau has nearly the same height in
all three cases, although the couplings differ by a factor of two. At
the same time, the spectra of Fig.~\ref{fig1}(a) split into a lower band
around the Fermi level and an upper band near the multiplet splitting
$\Delta_{\rm ex}=J(S+1/2)$.

\begin{figure}
    \centering
    \includegraphics[width=\columnwidth]{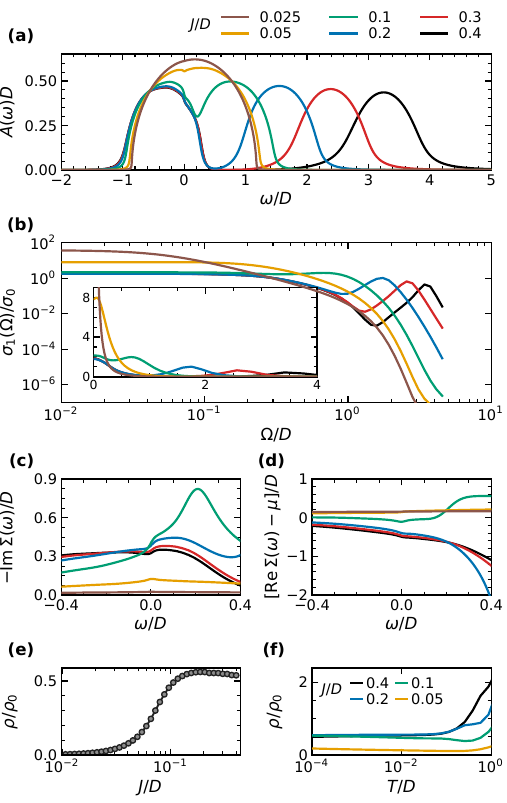}
    \caption{
    Resistivity saturation for $S=8$ at filling $n=0.8$.
    Panels (a--e) use $T/D=10^{-2}$.
    (a) Spectral function $A(\omega)D$.
    (b) Optical conductivity $\sigma_1(\Omega)/\sigma_0$, with a linear-scale
    inset.
    (c) Damping $-\mathrm{Im}\,\Sigma(\omega)/D$.
    (d) Real part of the self-energy relative to the chemical potential,
    $[\mathrm{Re}\,\Sigma(\omega)-\mu]/D$.
    (e) Resistivity $\rho/\rho_0$ versus $J/D$ at fixed temperature.
    (f) Resistivity versus $T/D$ for the indicated couplings.
    }
    \label{fig:saturation}
\end{figure}

Figure~\ref{fig:saturation}(a) follows the spectrum at $T/D=10^{-2}$ as
$J$ increases. For $J/D\leq0.05$ there is a single band. At $J/D=0.1$ a
dip opens above the Fermi level, and for $J/D\geq0.2$ the upper band
separates and shifts to higher energy as $\Delta_{\rm ex}$ increases. The lower band, formed
by electrons whose spin is antiparallel to the local moment, then becomes
nearly independent of $J$. This is the double-exchange limit, with the sign of the exchange
reversed \cite{AndersonHasegawa1955,Furukawa1994}: for $JS\gg D$, an
electron keeps its spin antiparallel to the local moment, because the
parallel configuration costs $\Delta_{\rm ex}$. Projected onto these lower
multiplets, an electron hops between sites with an amplitude set by the
overlap of the local spin states, $\cos(\theta_{ij}/2)$ for classical
moments at relative angle $\theta_{ij}$. This amplitude is independent of
$J$ to leading order; corrections are of order $D^2/\Delta_{\rm ex}$
\cite{Masui2022}. In this limit, $J$ sets only the separation between the
exchange multiplets.

The optical conductivity, Fig.~\ref{fig:saturation}(b), separates
accordingly into two sectors, as in the double-exchange model
\cite{Chattopadhyay2000}. The intraband part, $\Omega\lesssim D$, also
becomes independent of $J$. The interband peak, from transitions
across the exchange splitting between the configurations with the
conduction-electron spin antiparallel and parallel to the local moment, keeps
moving with $\Delta_{\rm ex}$. In Fig.~\ref{fig3}(d), the satellite appears 
within 10\% of $J(S+1/2)$ for
$S=1$--$8$. The
self-energy also becomes independent of $J$, after subtracting $\mu$ from its real part
[Fig.~\ref{fig:saturation}(c,d)]. Both
$\operatorname{Re}\Sigma$ and $\mu$ follow the energy of the lower
multiplet, $E_-=-J(S+1)/2$, and the shift cancels in
$\omega+\mu-\operatorname{Re}\Sigma$ at fixed filling
\cite{Furukawa1995}. The large damping reflects spin-disorder scattering: in
the paramagnet the lower-band electrons move through a background of
randomly oriented, nearly static moments. On these scales the system
therefore behaves as a disordered metal \cite{Furukawa1994}. Accordingly, the
low-frequency optical conductivity is a broad Drude peak that falls to
half its dc value only at $\Omega\approx0.32D$, a width set by the
damping rather than by $T$ [Fig.~\ref{fig:saturation}(b)].

Since the low-energy spectrum and self-energy saturate, so does the
resistivity. At $T/D=10^{-2}$, $\rho$ grows roughly as $J^2$ at small
coupling, as expected for Born spin-disorder scattering
\cite{Kasuya1956,deGennes1958}, and saturates at $\approx 0.54\rho_0$ for large $J$
[Fig.~\ref{fig:saturation}(e)]. As a function of temperature
[Fig.~\ref{fig:saturation}(f)], $\rho$ for $J/D=0.2$ and $0.4$ stays
within about 10\% of its value at $T/D=10^{-2}$ over several 
decades. It rises only as $T$
approaches the bandwidth (Sec.~\ref{sec:ultrahigh}). For $J/D\leq0.1$,
by contrast, the resistivity still increases on cooling, as in the
weak-coupling Kondo regime.

At fixed filling, the plateau approaches a coupling-independent classical value. For classical spins and $J\to\infty$,
the paramagnetic DMFT problem on the Bethe lattice can be solved in
closed form \cite{Furukawa1994}. For a classical spin along a unit
vector $\mathbf n$, the electron feels a static local field
$(JS/2)\,\mathbf n\cdot\boldsymbol\sigma$, with levels $\mp JS/2$ for spin
antiparallel and parallel to $\mathbf n$, so the impurity problem is
noninteracting. In the paramagnet, $\mathbf n$ is uniformly distributed, and
averaging over $\mathbf n$ gives each level weight $1/2$ in each spin component,
\begin{equation}
  G(z)=\frac12\sum_{s=\pm}\frac{1}{z+\mu-\Delta(z)+sJS/2},
  \label{eq:classical_spin_average}
\end{equation}
with $\Delta=D^2G/4$ from Eq.~\eqref{eq:dmft_bethe}. For $JS\gg D$, the
parallel level contributes only at order $D/JS$ within the lower band.
Keeping the antiparallel level, at $E_-^{\rm cl}=-JS/2$, gives
$G=1/[2(\zeta-D^2G/4)]$, where $\zeta=z+\mu-E_-^{\rm cl}$. This gives
\begin{equation}
\begin{aligned}
  G(z)&=\frac{2}{D^2}\Bigl[\zeta-\sqrt{\zeta^2-D^2/2}\Bigr],\\
\end{aligned}
\label{eq:de_limit}
\end{equation}
The lower band is thus a semicircle of weight $1/2$ and
half-width $D/\sqrt2$.
Equation~\eqref{eq:de_limit} gives the coherent-potential Green function of a
binary alloy with equal concentrations in the split-band limit, which is
exact in infinite dimensions: for classical spins, spin disorder acts as
substitutional disorder, as in the paramagnetic metal of the
Falicov--Kimball model \cite{Furukawa1994,Freericks2003}. At
$n=0.8$ and $T/D=10^{-2}$,
Eqs.~\eqref{eq:transport_spectrum}--\eqref{eq:dc_transport} give the
double-exchange (DE) resistivity $\rho_{\rm DE}=0.53\rho_0$, with $-\operatorname{Im}\Sigma(0)=0.31D$
and $A(0)D=0.39$, close to the $S=8$ results.
$J$ only confines electrons to the lower band, leaving the plateau
independent of $J$ \cite{Millis1995}. For quantum spins, the paramagnetic
double-exchange resistivity was computed in a many-body coherent-potential
approximation \cite{Edwards1999}.

Since the plateau is a double-exchange effect, the relevant parameter
is the classical exchange field $JS$, which sets
$\Delta_{\rm ex}\simeq JS$ relative to the bandwidth; quantum
corrections are of relative order $1/S$. Once $JS\gtrsim D$, the
low-energy physics no longer depends on $J$ and $S$ separately to leading
order in $1/S$, so for large spins the plateau is reached already at modest $J$. In
Fig.~\ref{fig1}(b,c), the curves for $S=4$ at $J/D=0.4$ and for $S=8$
at $J/D=0.2$, both with $JS=1.6D$, agree to within 3\% for
$10^{-2}\leq T/D\leq10^{-1}$, and all curves with $JS\geq0.8D$ lie
between $0.53\rho_0$ and $0.56\rho_0$ at $T/D=10^{-2}$.

\subsection{Filling dependence}
\label{sec:filling}

\begin{figure}
    \centering
    \includegraphics[width=\columnwidth]{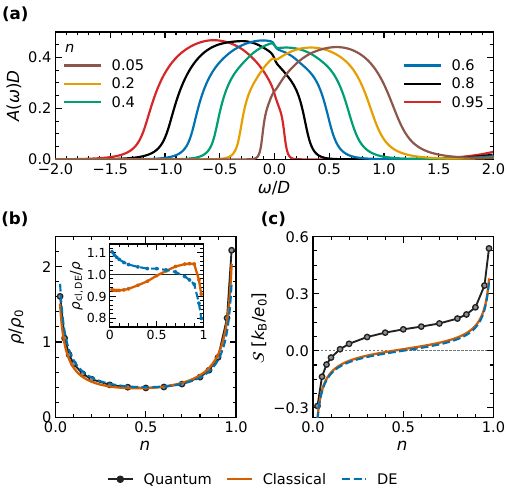}
    \caption{
    Filling dependence of resistivity saturation for $S=8$ at $J/D=0.4$
    and $T/D=10^{-2}$.
    (a) Spectral function $A(\omega)D$; colors indicate the band filling $n$.
    (b,c) Resistivity and Seebeck coefficient versus $n$.
    Symbols are quantum-spin results; orange solid curves are classical-spin
    DMFT at the same $JS$ and $T$, and blue dashed curves are the projected
    double-exchange (DE) limit. The inset in (b) shows the classical and DE
    resistivities, $\rho_{\rm cl}$ and $\rho_{\rm DE}$, divided by the
    quantum-spin resistivity $\rho$.
    }
    \label{fig:filling}
\end{figure}

Figure~\ref{fig:filling} tests this picture across the whole lower band,
for $S=8$, $J/D=0.4$, and $T/D=10^{-2}$. As $n$ increases from $0.05$
to $0.95$, the Fermi level moves from the bottom to the top edge of the
lower band [Fig.~\ref{fig:filling}(a)]. The
band retains its general shape and full width of about $1.4D$, close to the
$\sqrt2D$ of Eq.~\eqref{eq:de_limit}, and $A(0)$ agrees with the
double-exchange value to within 1\% for $0.05\leq n\leq0.8$. The
remaining differences include finite-exchange and quantum-spin corrections. Adding an
electron of either spin to an empty site gives $2(2S+1)$ states, of which
only $2S$ belong to the lower multiplet. Averaged over the local spin states,
electron addition therefore contributes weight $S/(2S+1)$ per spin projection
to the lower band, rather than $1/2$. Removing the electron from a
lower-multiplet site always leaves an empty site and contributes weight $1/2$.
With empty and singly occupied sites in proportions $1-n$ and $n$, the lower
band therefore holds $n/2+(1-n)S/(2S+1)$ states per spin projection
instead of $1/2$. 

The relative imbalance of the conditional removal and addition weights is
\begin{equation}
\eta_R\equiv
\frac{\tfrac12-\tfrac{S}{2S+1}}{\tfrac12+\tfrac{S}{2S+1}}
=\frac{1}{4S+1}.
\label{eq:removal_addition_imbalance}
\end{equation}
Figure~\ref{fig:filling}(a) also shows a thermally rounded downward step
at the Fermi level, superposed on the broad lower band. As a working
hypothesis in the plateau regime, away from band edges, suppose the removal
and addition spectra share a common envelope $A_{\rm sm}(\omega)$ that
varies slowly across the thermal window. Assigning their relative amplitudes
from the conditional weights motivates the local ansatz
\begin{equation}
A(\omega)\simeq A_{\rm sm}(\omega)
\left[1-\eta_R\tanh\frac{\omega}{2T}\right].
\label{eq:spectral_step_hypothesis}
\end{equation}

The resistivity
[Fig.~\ref{fig:filling}(b)] is lowest, $0.39\rho_0$, at
$n=1/2$ and rises steeply as the lower band empties or fills up. The
DE resistivity $\rho_{\rm DE}(n)$, obtained from
Eq.~\eqref{eq:de_limit} at each filling \cite{Furukawa1994}, depends only
on the filling of the lower band. It is symmetric about
$n=1/2$, where it has its minimum, $0.40\rho_0$, and diverges as $n\to0$
or $n\to1$. For the $S=8$ resistivity $\rho$, the ratio $\rho_{\rm DE}/\rho$
decreases from $1.09$ at $n=0.05$ to $0.96$ at $n=0.9$ and $0.80$ at
$n=0.975$ [inset of Fig.~\ref{fig:filling}(b)].
To separate finite-exchange from quantum-spin effects,
Fig.~\ref{fig:filling}(b) also shows a classical-spin reference:
Eq.~\eqref{eq:classical_spin_average} with $\Delta=D^2G/4$, solved at the
same $JS/2=1.6D$ and $T/D=10^{-2}$ without projecting out the upper
exchange band. This classical resistivity, $\rho_{\rm cl}$, differs from
the $S=8$ result only by treating the spins classically, and from
$\rho_{\rm DE}$ only by retaining the upper band. In the inset,
$\rho_{\rm cl}/\rho$ therefore measures the quantum-spin correction, and the
separation between $\rho_{\rm cl}/\rho$ and $\rho_{\rm DE}/\rho$ measures
the finite-exchange correction.

\subsection{Quantum thermopower beyond the Kelvin approximation}
\label{sec:quantum_thermopower}

The Seebeck coefficient [Fig.~\ref{fig:filling}(c)] is negative at low
filling, changes sign at $n\approx0.14$, grows slowly and almost linearly
for $0.3\lesssim n\lesssim0.7$, and rises steeply as the lower band fills
up. In the double-exchange limit, $\mathcal K(\omega)$ of
Eq.~\eqref{eq:transport_moments} varies smoothly across the thermal window,
and expanding it to linear order in $\omega$ gives
$\mathcal S/(k_B/e_0)\simeq-(\pi^2/3)\,T\,\mathcal K'(0)/\mathcal K(0)$.
Because the lower band is
symmetric about its center, the DE thermopower is antisymmetric about $n=1/2$,
electron-like below and hole-like above. The $S=8$ results instead lie above this DE
thermopower, $\mathcal S_{\rm DE}$, by an almost constant $0.10$--$0.11\,k_B/e_0$ for
$0.2\leq n\leq0.9$, shifting the zero to $n\approx0.14$.
Thus quantum spins reverse the sign of thermopower
over a broad filling interval even while resistivity remains close
to its classical value.

Using the finite-$JS$ classical reference at matched filling and
temperature \cite{Furukawa1994}, we define the quantum excess of the Kubo
thermopower, Eq.~\eqref{eq:dc_transport}, as
$\Delta\mathcal S=\mathcal S_q-\mathcal S_{\rm cl}$, where $q$ and
$\mathrm{cl}$ denote the quantum and classical results. At $n=0.8$,
$J/D=0.4$, and $T/D=0.01$, the resistivity ratios
$\rho_q/\rho_{\rm cl}$ are $0.99$ and $0.96$ for $S=4$ and $8$,
respectively, yet the thermopower excesses remain $0.213$ and
$0.101\,k_B/e_0$. The excess thus survives finite-$JS$ corrections.

The Fermi-level step in Fig.~\ref{fig:filling}(a) is accompanied by an
asymmetry of the transport kernel. The quantum thermopower excess is
determined by the difference between the normalized odd parts of the quantum
and classical kernels. With
$\mathcal K_\pm(\omega)=[\mathcal K(\omega)\pm\mathcal K(-\omega)]/2$,
only $\mathcal K_+$ contributes to $L_0$ and only $\mathcal K_-$ to
$L_1$. To compare thermopowers with different conductivities, define
$\widehat{\mathcal K}=\mathcal K/L_0$ separately for each solution.
With $x=\omega/T$ and $w(x)=[4\cosh^2(x/2)]^{-1}$, the excess is exactly
\begin{equation}
\begin{aligned}
\mathcal I_{\Delta\mathcal S}(x)
 &=-2xw(x)\bigl[\widehat{\mathcal K}_{q,-}(Tx)
                  -\widehat{\mathcal K}_{{\rm cl},-}(Tx)\bigr],\\
\frac{\Delta\mathcal S}{k_B/e_0}
 &=\int_0^\infty dx\,\mathcal I_{\Delta\mathcal S}(x).
\end{aligned}
\label{eq:quantum_excess}
\end{equation}
For $S=8$, $n=0.8$, and $J/D=0.4$, the odd-kernel difference is a
thermally smeared step at the Fermi level,
$\widehat{\mathcal K}_{q,-}(Tx)-\widehat{\mathcal K}_{{\rm cl},-}(Tx)\simeq-\delta\tanh(x/2)$,
to within 3\% of its dimensionless amplitude $\delta$, which is half the
full step height. Because
$\int_0^\infty dx\,2xw(x)\tanh(x/2)=1$, the excess is
$\Delta\mathcal S\simeq\delta\,k_B/e_0$, and the integrand peaks at
$\omega\simeq2T$. The amplitude, and with it the excess, changes little
with temperature: $\Delta\mathcal S$ increases only from $0.093$ to
$0.101\,k_B/e_0$ between $T/D\simeq10^{-4}$ and $10^{-2}$.

The spectral working hypothesis, Eq.~\eqref{eq:spectral_step_hypothesis},
motivates an approximately proportional transport-step amplitude,
$\delta\simeq C\eta_R$, at leading order in the small imbalance.
The dimensionless coefficient $C$ reflects transport weighting and can
depend on filling, exchange, and temperature. At $n=0.8$ and $J/D=0.4$,
the normalized excess $\Delta\mathcal S/[(k_B/e_0)\eta_R]$ is $3.06$ and
$3.07$ for $S=4$ and $8$ at $T/D\simeq1.16\times10^{-4}$; at
$T/D=10^{-2}$ the values are $3.62$ and $3.34$. The numerical spin
dependence of the thermopower excess is consistent with this interpretation,
although it does not establish the assumed spectral form or a universal
proportionality coefficient.

\begin{figure}[t]
    \centering
    \includegraphics[width=\columnwidth]{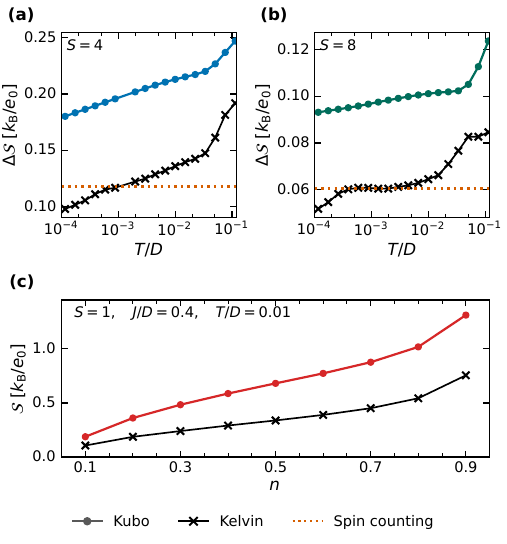}
    \caption{
    (a,b) Kubo and Kelvin thermopower excesses at $J/D=0.4$ and $n=0.8$ for
    $S=4$ and $8$, relative to classical spins at the same $JS$, $n$, and $T$.
    Dotted lines show the spin-counting estimate $\ln[(2S+1)/(2S)]$.
    (c) Total Kubo and Kelvin thermopower for $S=1$ at $J/D=0.4$ and
    $T/D=0.01$ versus filling, with the symbols of (a,b).
    }
    \label{fig:quantum_excess}
\end{figure}

To estimate the excess thermodynamically, note that thermopower is more
sensitive than resistivity to local spin and orbital degeneracies in correlated hopping models \cite{Koshibae2001,Wang2003}. 
Local degeneracy counting, as in generalized Heikes formulas, gives a
spin-entropy estimate \cite{Chaikin1976,Koshibae2000}.
For $T\ll\Delta_{\rm ex}$, adding an electron to the lower band
turns a site of spin $S$ into one of spin $S-1/2$ and lowers the spin
entropy by $k_B\ln[(2S+1)/2S]$. Using the Kelvin approximation
$\mathcal S\simeq-(1/e_0)(\partial s/\partial n)_T$, with $s$ the entropy
per site \cite{Peterson2010,Mravlje2016,Shastry2013}, gives a positive,
filling-independent contribution $(k_B/e_0)\ln[(2S+1)/2S]$, which is
$0.06\,k_B/e_0$ for $S=8$ and vanishes for classical spins. The estimate
reproduces the sign, filling independence, and $S$ dependence,
but accounts for only about two-thirds of the magnitude. At $n=0.8$, $J/D=0.4$, and
$T/D=10^{-4}$, where the band contribution is negligible, the
thermopower in Fig.~\ref{fig1}(d) is $0.182\,k_B/e_0$ for $S=4$ and
$0.094\,k_B/e_0$ for $S=8$. Their ratio, $1.94$, equals that of the
estimates, $\ln(9/8)/\ln(17/16)=1.94$: the counting captures the
$S$ dependence, while both values fall short by the same factor.

We test the full Kelvin approximation independently of multiplet counting using
$\mathcal S_K/(k_B/e_0)=(\partial\mu/\partial T)_n$
\cite{Peterson2010,Mravlje2016}. Figure~\ref{fig:quantum_excess}(a,b) compares this Kubo excess,
$\Delta\mathcal S$, with the Kelvin excess $\Delta\mathcal S_K=\mathcal S_{K,q}-\mathcal S_{K,{\rm cl}}$.
At $T/D=0.01$, the latter is $0.136$ for $S=4$ and $0.0645$ for $S=8$,
in units of $k_B/e_0$, about $64\%$ of the respective Kubo excesses.
Spin entropy therefore explains the sign and spin trend,
but neither local counting nor the full Kelvin approximation captures the
complete transport-weighted asymmetry. Sensitivity to asymmetric
scattering also underlies the thermopower of skewed non-Fermi liquids
\cite{GeorgesMravlje2021} and its application to cuprates
\cite{Gourgout2022}.

The Kelvin underestimate also extends to smaller spins and other fillings:
for $S=1$, $J/D=0.4$, and $T/D=0.01$, it persists throughout
$0.1\leq n\leq0.9$ [Fig.~\ref{fig:quantum_excess}(c)]. Throughout this
interval, the Kelvin value is only 50--58\% of the Kubo thermopower.
In the Hund metal Sr$_2$RuO$_4$, by contrast, thermopower and its Kelvin
estimate both track the spin entropy released after orbital fluctuations
are quenched \cite{Mravlje2016}. The Kelvin approximation is likewise useful
for incoherent transport in the strongly correlated Hubbard model
\cite{Arsenault2013Thermopower}, and it is exact for lattices of
Sachdev--Ye--Kitaev islands \cite{Davison2017}. Since both the Kubo bubble and
$(\partial\mu/\partial T)_n$ are exact within DMFT, the shortfall found here is
intrinsic to exchange scattering from the local moments rather than an artifact
of the calculation.

\subsection{Low-temperature breakdown of the plateau}
\label{sec:classical_scaling}

The plateau is not a ground-state property. Even within the computed range,
the damping at the Fermi level decreases slowly: for $S=8$ at $J/D=0.4$ it
drops from $0.32D$ at $T/D=10^{-2}$ to $0.22D$ at $10^{-6}$. At the lowest temperatures,
curves with equal $JS$ separate: at $T/D=10^{-5}$ the $S=4$ resistivity
has dropped to $0.43\rho_0$, while the $S=8$ value is still
$0.53\rho_0$. This decrease is a quantum effect of the residual spins,
which form the underscreened singular Fermi liquid of
Sec.~\ref{sec:lowT}; its rate scales as $1/S_*$
(Appendix~\ref{app:larger_spins}). Even with a local exponent of $0.02$,
halving the resistivity would require cooling by about 15 decades.

This slow crossover makes the classical limit nonuniform in temperature.
At fixed nonzero $JS$ and metallic filling, within the homogeneous
paramagnetic solution,
\begin{equation}
\begin{aligned}
 \lim_{S\to\infty}\lim_{T\to0}\rho(T,S)&=0,\\
 \lim_{T\to0}\lim_{S\to\infty}\rho(T,S)&=\rho_{\rm cl}(0)>0.
\end{aligned}
\label{eq:noncommuting_spin_limits}
\end{equation}
The first relation follows from the finite-spin singular-Fermi-liquid
asymptote; the second from classical spin-disorder scattering.

\section{High-temperature regime and incoherent Kondo scattering}
\label{sec:highT}

\subsection{Running coupling and resistivity fits}
\label{sec:kondo_fits}

\begin{figure}[t]
    \centering
    \includegraphics[width=\columnwidth]{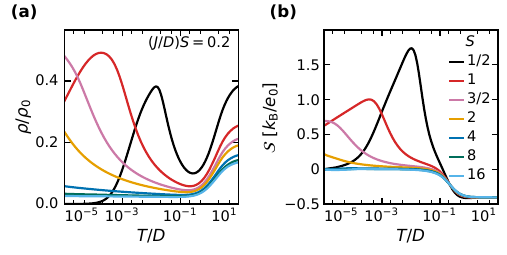}
    \caption{
    Transport at fixed $JS/D=0.2$ and filling $n=0.8$ for
    $S=1/2$, $1$, $3/2$, $2$, $4$, $8$, and $16$.
    (a) Normalized resistivity $\rho/\rho_0$.
    (b) Seebeck coefficient in units of $k_{\mathrm B}/e_0$.
    }
    \label{fig:fixedJS_fits}
    \label{fig:fixedJS_transport}
\end{figure}

For low $J$, above the Kondo scale, the local moments are no longer screened. For
$T_K\ll T\ll D$ each spin fluctuates freely and independently of its
neighbors. Electrons scatter incoherently from these spins as from one Kondo
impurity per site. The self-energy is small,
and the DMFT bath is close to the bare band. To second order in $J$,
exchange scattering gives
\begin{equation}
\begin{aligned}
  -\operatorname{Im}\Sigma&=\frac{\pi}{4}\nu_FJ^2S(S+1),\\
  \frac{\rho}{\rho_0}&=\frac{-2\operatorname{Im}\Sigma}{\pi\Phi_F}
  =A_{\rm th}\,g^2,
\end{aligned}
\label{eq:born_spin_disorder}
\end{equation}
with $g=\nu_FJ$ and $A_{\rm th}=S(S+1)/(2\nu_F\Phi_F)$. Here $\nu_F$ and
$\Phi_F$ are the bare density of states and the transport
function, Eq.~\eqref{eq:transport_function}, at the noninteracting Fermi
level. This is the spin-disorder resistivity of independent local
moments \cite{Kasuya1956,deGennes1958}. At fixed $JS=0.2D$, where $T_K$
lies below the temperatures shown, the large-spin resistivities in
Fig.~\ref{fig:fixedJS_transport} approach this Born value,
$\rho\propto(JS)^2$, as $S$ increases. At third order, the commutator
of the spin operators produces Kondo's logarithm \cite{Kondo1964}.
Resumming the leading logarithms replaces $g$ by the running coupling
$g(T)$, evaluated at the temperature of interest. For antiferromagnetic exchange it
grows on cooling, $g(T)\simeq1/\ln(T/T_K)$ with $T_K=De^{-1/g}$ at this
order, so the resistivity rises logarithmically as $T$ approaches $T_K$
from above. For quantitative fits we integrate the two-loop beta
function \cite{AbrikosovMigdal1970,Fowler1971} of
Appendix~\ref{app:two_loop}, $\beta(g)=dg/d\ln E=-g^2+g^3/2$,
for $g>0$:
\begingroup
\displaywidowpenalty=10000
\begin{equation}
\begin{gathered}
  \frac{\rho(T)}{\rho_0}=\frac{A}{y(T)^2},\\[2pt]
  y(T)+\frac12\ln\!\bigl[y(T)-\tfrac12\bigr]
    =\ln\frac{T}{T_0},
\end{gathered}
\label{eq:kondo_fit}
\end{equation}
\endgroup
with $y=1/g(T)>1/2$. Only the amplitude $A$ and the scale $T_0$ are
fitted; Eq.~\eqref{eq:born_spin_disorder} predicts $A=A_{\rm th}$.
The spin sets the strength of the scattering through $S(S+1)$, while
the coupling alone controls the RG equations.

\begin{figure}[t]
    \centering
    \includegraphics[width=3.375in]{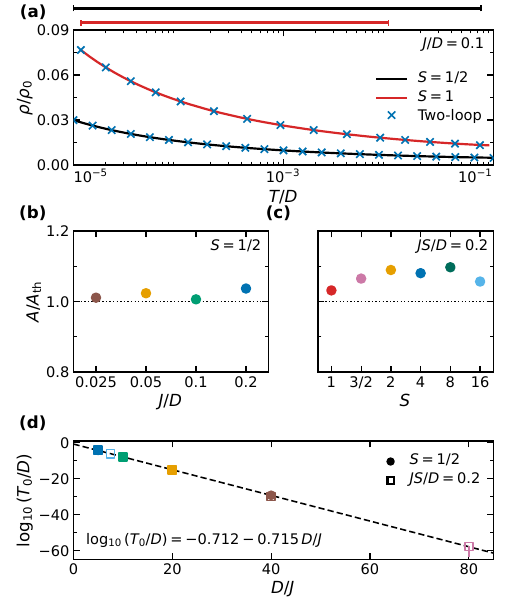}
    \caption{
    Two-loop Kondo fits at filling $n=0.8$. (a) DMFT resistivity (solid)
    and model (blue crosses). Horizontal bars above panel (a) mark fitted intervals
    (black: $S=1/2$; red: $S=1$). (b,c) Amplitudes relative to
    $A_{\rm th}=S(S+1)/(2\nu_F\Phi_F)$.
    (d) Integration scales $T_0$ vs. $J$.
    }
    \label{fig:kondo_scaling}
\end{figure}

Figure~\ref{fig:kondo_scaling} compares Eq.~\eqref{eq:kondo_fit} with
the DMFT resistivity. Both spins give the same scale,
$T_0=1.4\times10^{-8}D$, and amplitudes $A/A_{\rm th}=1.01$ and $1.02$.
Panels (b) and (c) collect the
amplitudes for $S=1/2$ at $0.025\le J/D\le0.2$ and for $S=1$--$16$ at
fixed $JS=0.2D$. 
The scales in panel (d) depend on $J$ but hardly on $S$, as
expected, since the spin does not enter the coefficients of the
weak-coupling flow. For $S=1/2$, $\log(T_0/D)$
is approximately linear in $D/J$, with a fitted slope within 2\% of that obtained from
$T_0\propto\sqrt g\,e^{-1/g}$.

\subsection{Relation to the low-temperature singular Fermi liquid}
\label{sec:kondo_duality}

\begin{table}[b]
  \caption{The two weak-coupling regimes of the underscreened Kondo
  lattice. Both follow the same flow, $dg/d\ln E=-g^2+g^3/2$, on
  opposite sides of $g=0$. Here $c$ is the dimensionless exchange field,
  $|g(T)|$ times the scattering spin.}
  \label{tab:duality}
  \begin{ruledtabular}
  \begin{tabular}{lcc}
    & $T\gg T_K$ & $T\ll T_K$\\
    \hline
    Scattering spin & bare, $S$ & residual, $S_*=S-1/2$\\
    Exchange & antiferromagnetic & ferromagnetic\\
    Coupling $g(T)$ & $\simeq1/\ln(T/T_K)$ & $\simeq-1/\ln(T_{\rm s}/T)$\\
    On cooling & $|g(T)|$ grows & $|g(T)|$ decreases\\
    Resistivity & $\propto S(S+1)\,g(T)^2$ & $\propto S_*(S_*+1)\,g(T)^2$\\
    Two-loop form & Eq.~\eqref{eq:kondo_fit} & Eq.~\eqref{eq:sfl_rho_fit}\\
    $d\ln\rho/d\ln T$ & $-2c/S$ & $+2c/S_*$\\
  \end{tabular}
  \end{ruledtabular}
\end{table}

The weak-coupling regime above $T_K$ is the mirror image of the singular
Fermi liquid below it (Sec.~\ref{sec:lowT}); Table~\ref{tab:duality}
lists the correspondences. In both regions, electrons scatter incoherently from free, randomly oriented spins,
giving the Born spin-disorder resistivity with a logarithmically running
weak coupling. They differ in
the scattering spin and the sign of the exchange coupling
\cite{Nozieres1980,Coleman2003,Mehta2005}. Above $T_K$, the bare spin
$S$ couples antiferromagnetically; the coupling is marginally relevant
and grows on cooling. Below $T_K$, the screening channel has absorbed
half a unit of spin, and the residual spin $S_*=S-1/2$ couples
ferromagnetically; the coupling is marginally irrelevant and decreases
on cooling \cite{Koller2005,CabreraCano2013}. The same beta function
governs both flows, on the two sides of $g=0$. Equations
\eqref{eq:kondo_fit} and \eqref{eq:sfl_rho_fit} map onto each other
under $g\to-g$, with $T/T_0$ replaced by $T_\rho/T$.
The resistivity therefore falls off as an inverse squared logarithm on
both sides of $T_K$: as $S(S+1)/\ln^2(T/T_0)$ above and as
$S_*(S_*+1)/\ln^2(T_\rho/T)$ below. The first corrections,
$\pm\ln\ell/\ell^3$ with $\ell$ the large logarithm, have opposite signs
because the two-loop term of $\beta(g)$ is odd in $g$.
In between, the coupling is strong and the resistivity passes through
its maximum at $T \eqsim T_K$.

The duality extends to the dependence on spin. On both sides, the
strength of the scattering is set by the dimensionless exchange field,
$c=g(T)S$ above $T_K$ and $c=|g(T)|S_*$ below, whereas the logarithms come
from the spin commutator, which is linear in the spin
(Appendix~\ref{app:larger_spins}). At fixed exchange field, the leading
logarithmic flow rate scales as $1/S$ above $T_K$ and $1/S_*$ below.

\section{Ultrahigh-temperature regime and $T$-linear resistivity}
\label{sec:ultrahigh}

\subsection{Charge susceptibility and linear resistivity}
\label{sec:charge_linear}

\begin{figure}[!t]
    \centering
    \includegraphics[width=\columnwidth]{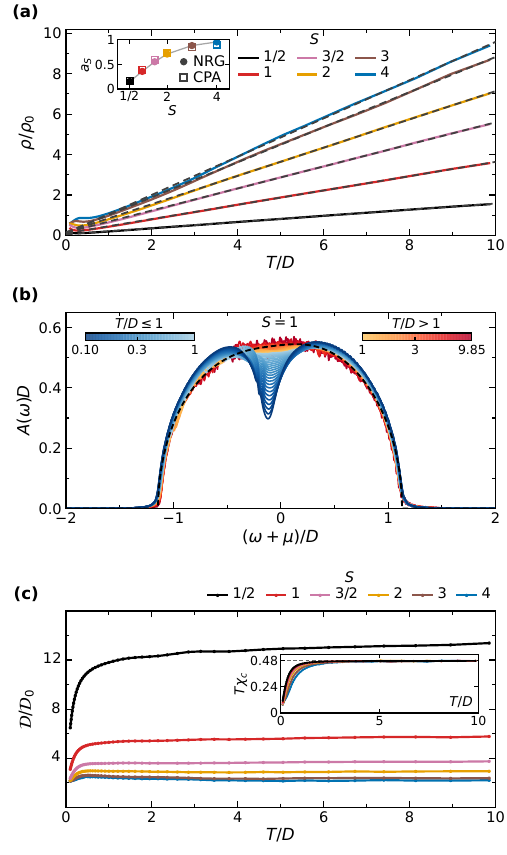}
    \caption{
    Ultrahigh-temperature transport at $n=0.8$ and $J/D=0.4$.
    (a) $\rho/\rho_0$ for a range of $S$.
    Gray dashed lines are linear fits using $3\le T/D\le10$.
    Inset slopes $a_S$ come from fits $\rho/\rho_0=a_S(T/D)+b_S$
    over $1\le T/D\le10$ (filled circles); open squares show the
    infinite-temperature CPA prediction at the same $J$ and $n$.
    (b) Local $S=1$ spectrum $A(\omega)D$ versus $(\omega+\mu)/D$;
    the $T>D$ curves nearly overlap. The dashed curve is the
    infinite-temperature CPA spectrum.
    (c) Effective diffusion constant $\mathcal D/\mathcal D_0=(\sigmadc/\sigma_0)/(D\chi_c)$,
    with $\mathcal D_0=\sigma_0Dv_0/e_0^2$ and site volume $v_0$.
    Inset: $T\chi_c$; dashed line: infinite-temperature limit.
    }
    \label{fig:high_temperature}
\end{figure}

Figure~\ref{fig:high_temperature} examines the renewed increase of resistivity
at temperatures comparable to and above the conduction bandwidth. 
Panel (a) shows linear temperature dependence, reaching resistivities well above $\rho_0$
without saturation; the inset shows the spin dependence of the slopes.
Thus, although the finite
local-spin Hilbert space limits the growth of spin disorder, it does not
impose an upper bound on the resistivity.

Panel (b) illustrates the accompanying spectral evolution for $S=1$. The
low-temperature depletion fills in on heating, and the spectra nearly
coincide for $T>D$. Their width remains set by the band and exchange energy
scales. The energy axis $E=\omega+\mu$ removes the chemical-potential shift:
away from half filling, the spectrum in that limit is a fixed function of $E$,
not of the frequency measured from $\mu$. Resistivity therefore continues to increase
while the local spectrum changes little.

At fixed filling, the high-temperature chemical potential and Fermi-window
weight satisfy
\begin{equation}
  \frac{\mu}{T}\longrightarrow\ln\frac{n}{2-n},
  \qquad
  -f'(E-\mu)\sim\frac{n(2-n)}{4T}
  \label{eq:highT_fermi_window}
\end{equation}
for finite $E$. Since the transport kernel of Eq.~\eqref{eq:transport_moments}
approaches a temperature-independent function of $E$ with a finite integral, the $1/T$ thermal weight gives
$\sigmadc\propto T^{-1}$ and $\rho_\mathrm{dc}\propto T$. The
near overlap of the spectra in panel (b) is consistent with this mechanism.
Thermal weighting also produces $T$-linear transport for incoherent
carriers coupled to classical local-moment fields \cite{Ciuchi2023} and in
the $t$--$J$ model \cite{Fratini2026}. Such linearity alone does not establish
Planckian relaxation, $\hbar/\tau\sim T$, since conductivity also depends on
thermodynamic and optical weights \cite{Hartnoll2022}. The unbounded linear increase
does not contradict the saturation of weak-coupling scattering from
nearly classical local moments at temperatures above the intermoment exchange scale
but below the Fermi energy \cite{McRoberts2023}: our linear regime lies
above the electronic bandwidth. Nor does the resistivity have to saturate at the
Mott--Ioffe--Regel scale, which is of order $\rho_0$: such saturation is
mechanism dependent \cite{Gunnarsson2003}.

We separate contributions using the Nernst--Einstein
relation~\mbox{\cite{Kokalj2017,Perepelitsky2016,Hartnoll2014}}
\begin{equation}
  \sigmadc=\frac{e_0^2}{v_0}\chi_c\mathcal D,
  \qquad
  \chi_c=\left(\frac{\partial n}{\partial\mu}\right)_T,
  \label{eq:einstein_decomposition}
\end{equation}
where $n$ is the conduction-electron occupancy per site and $v_0$ is the
site volume. We estimate $\chi_c$ from the computed occupancies and chemical
potentials of neighboring DMFT solutions with target fillings $0.79$ and
$0.8$. Panel (c) shows the resulting effective diffusion constant $\mathcal D$,\footnote{This effective diffusion constant need not coincide with a diffusion eigenvalue when charge and heat transport mix \cite{Ulaga2023}.}
which becomes only weakly temperature dependent at high $T$.

The inset identifies the origin of the remaining temperature dependence of
the conductivity. For $T\gg D,J(S+1/2)$, the localized-spin multiplicity
cancels from the charge-state probabilities, giving
\begin{equation}
  \chi_c\sim\frac{n(2-n)}{2T}.
  \label{eq:highT_charge_susceptibility}
\end{equation}
Thus $T\chi_c\to0.48$ at $n=0.8$, independently of $S$. All curves approach this
common limit. Thermopower likewise approaches the spin-independent value
$\mathcal S/(k_B/e_0)\to\ln[n/(2-n)]$, which gives $-0.405$ at $n=0.8$, consistent
with Fig.~\ref{fig1}(d,e). 

The transport kernel at high temperature depends on the exchange transitions, retaining microscopic spin
information absent from the charge-thermodynamic quantities in that limit.
The high-temperature increase of resistivity is therefore
primarily associated with the decreasing charge susceptibility, whereas
its spin-dependent slope reflects the different values of $\mathcal D$
in that limit.

This separation closely parallels our high-temperature expansion and DMFT
results for the Hubbard model \cite{Perepelitsky2016}. Kokalj
likewise found that, in the lightly doped two-dimensional Hubbard model,
the strong temperature and doping dependence of resistivity largely
follows the inverse charge susceptibility, with a comparatively weak
variation of the diffusion constant \cite{Kokalj2017}. The
charge-state counting differs here: the Kondo lattice permits double
occupancy, giving $T\chi_c\to n(2-n)/2$, whereas the $U=\infty$ Hubbard
model has $T\chi_c\to n(1-n)$.

\subsection{Exchange multiplets and the spin-dependent slope}
\label{sec:spin_slope}

The spin-dependent coefficient can be understood using the many-body
coherent-potential approximation (CPA) of Ref.~\cite{Edwards1999},
specialized to the infinite-temperature atomic weights and
antiferromagnetic exchange. The atomic electron-addition and
removal transitions retain both exchange multiplets and all charge sectors.
Embedding these transitions self-consistently in the semicircular band
gives a limiting transport kernel and hence a prediction
$a_S^{\rm CPA}$ for the infinite-temperature linear slope
(Appendix~\ref{app:highT_cpa}). At weak exchange,
$a_S^{\rm CPA}\propto (J/D)^2S(S+1)$; multiple scattering and redistribution of
spectral weight between the exchange bands suppress this rapid growth
at larger spins. For the parameters of Fig.~\ref{fig:high_temperature},
the CPA slopes, shown as open squares in the inset, differ from the DMFT
slopes fitted over $1\le T/D\le10$ by at most $8.2\%$.

The CPA spectrum itself is close to the numerical one: for $S=1$, the
dashed curve in Fig.~\ref{fig:high_temperature}(b) differs from the
$T>D$ spectra by 2--4\% in integrated absolute value.

\section{Conclusion}

Within the paramagnetic single-channel Kondo lattice, increasing $S$ at fixed
$JS$ takes transport from quantum Kondo behavior to classical spin-disorder
scattering. Incomplete
screening leaves residual moments whose dynamics control the slow
low-temperature evolution of transport. The apparent power laws at low
temperatures and the intermediate-temperature plateau span finite
temperature windows. For $S=1$, the drifting resistivity exponent is
consistent with the inverse-log-squared singular-Fermi-liquid asymptote and
its two-loop correction. For larger spins, the slower evolution makes the
asymptotic regime harder to resolve. Optics distinguishes full from
partial screening: for $S=1/2$, a narrow Drude peak is followed by the
Fermi-liquid foot, whereas the underscreened models exhibit a broader
response with non-Drude tails.

The ferromagnetic Kondo flow at low temperature and the antiferromagnetic
flow at high temperature share a weak-coupling description. Their
quantum corrections weaken with increasing $S$ at fixed $JS$, broadening the
classical transport window. For $JS\gtrsim D$, the plateau approaches the
double-exchange result while retaining an eventual logarithmic decrease at finite $S$.
Thermopower remains sensitive to quantum spin effects even where resistivity
is nearly classical. Multiplet counting captures the sign and spin trend of
the excess, but even the full Kelvin approximation captures only part of it;
for $S=1$ it gives about half of the thermopower. The odd transport kernel
thus retains a quantum contribution when its even part is already nearly classical.

Above the band and exchange scales, the charge susceptibility decreases as
$1/T$ while the effective diffusion constant is nearly constant. The resulting
$T$-linear resistivity does not require a scattering rate proportional to $T$.
The many-body CPA accounts for its spin-dependent coefficient by embedding
the atomic exchange multiplets self-consistently: thermodynamics fixes the
$T$ dependence, while microscopic dynamics set the resistivity scale.

These conclusions concern the homogeneous paramagnetic solution
within single-site DMFT. Magnetic ordering and nonlocal spin correlations
would modify or preempt its lowest-temperature behavior.

For a spin-1 impurity, hard-axis anisotropy smaller than $T_K$ induces
a second screening stage at an exponentially small scale
\cite{Zitko2008Anisotropy}, motivating a test of whether it cuts off the
slow lattice transport crossover.
Additional channels would permit comparison with fully screened and
overscreened lattices \cite{Jarrell1996}. Together with magnetic order,
these extensions would test which signatures survive splitting, screening,
or ordering of the residual moments.

\section{Data}

The DMFT/NRG code used in this work is available in Ref.~\cite{DMFTNRGKLM}.
The numerical data and code for reproducing the figures are available on
Zenodo~\cite{KLMData}.

\begin{acknowledgments}
We acknowledge the support of the Slovenian Research and Innovation Agency (ARIS) under P1-0416.
\end{acknowledgments}

\appendix
\section{Singular Fermi-liquid self-energy and dc transport}
\label{app:sfl_transport}

We derive the low-temperature transport law for the paramagnetic $S=1$ KLM
within DMFT, extending it to $S\geq3/2$ in the final subsection. We assume a
metallic self-consistent bath and $|\omega|,T\ll T_K$, the first-stage
Kondo screening scale.

\subsection{Residual exchange and the interaction self-energy}

The local Green function $G$ describes electron propagation on the impurity
orbital. The hybridization $\Delta$ describes propagation into the bath and
back, while the self-energy $\Sigma$ accounts for the local exchange
interaction. The Weiss propagator $\mathcal G_0^{-1}=z+\mu-\Delta$ retains
the bath but omits this interaction. Spin indices are suppressed in the
paramagnet.

After the first-stage Kondo screening of spin 1, the impurity's low-energy effective Hamiltonian has
the form \cite{Nozieres1980,Cragg1979,Koller2005,Mehta2005}
\begin{equation}
\begin{gathered}
  H_{\rm low}=H_*+J_{\rm r}\,\mathbf S_*\cdot\mathbf s_*(0),\\
  S_*=\frac12,\qquad J_{\rm r}<0.
\end{gathered}
\label{eq:sfl_effective_hamiltonian}
\end{equation}
Here $\mathbf S_*$ is the residual spin and $J_{\rm r}<0$ its
ferromagnetic exchange with the screening channel.
$H_*$ describes the screening-channel fermions with their fixed-point
elastic scattering. Their local spin density is
$\mathbf s_*(0)=\frac12\sum_{\alpha\beta}\psi_{*\alpha}^\dagger
\boldsymbol\sigma_{\alpha\beta}\psi_{*\beta}$, where $\psi_{*\alpha}$
annihilates a local screening-channel electron of spin $\alpha$.
The field $\psi_*$ includes fixed-point elastic scattering and is distinct
from the physical orbital $c$.
Let $\nu_*$ be its Fermi-level density of states per electron spin.
The dimensionless exchange $g=\nu_*J_{\rm r}$ has a unit one-loop
coefficient. The initial cutoff $D_*$ bounds the low-energy theory after
first-stage screening.
At this cutoff the coupling is
$g_0\equiv g(D_*)=\nu_*J_{\rm r}$. Eliminating electronic states between
$D_*$ and a lower scale $E$, with $\ell=\ln(D_*/E)$, gives
\begin{equation}
  \frac{dg}{d\ell}=g^2+O(g^3),\qquad
  g(E)=\frac{g_0}{1-g_0\ell}
  \label{eq:sfl_one_loop_flow}
\end{equation}
at one loop \cite{Anderson1970}. For $g_0<0$ this is Eq.~\eqref{eq:sfl_running}, with
$T_{\rm s}=D_*\exp(1/|g_0|)$. In the unpolarized residual-spin ensemble,
$\langle S_*^a\rangle=0$ and
$\langle S_*^aS_*^b\rangle=\delta_{ab}S_*(S_*+1)/3$,
where $a,b,c\in\{x,y,z\}$ and $\delta_{ab}$ is the Kronecker delta.
The first nonzero spin-averaged scattering rate is consequently
quadratic in $g$.

The impurity scattering amplitude must be distinguished from the
interaction self-energy in Eq.~\eqref{eq:dmft_lattice}. We define the local
impurity $T$-matrix $\mathcal T$ relative to the Weiss propagator by
$G=\mathcal G_0+\mathcal G_0\mathcal T\mathcal G_0$. Comparing with Dyson's equation,
$G=\mathcal G_0+\mathcal G_0\Sigma G$, gives $\mathcal T=\Sigma+\Sigma\mathcal G_0\mathcal T$.
Solving for $\Sigma$, we obtain
\begin{equation}
\begin{aligned}
  \Sigma^R&=\frac{\mathcal T^R}{1+\mathcal G_0^R\mathcal T^R},\\
  -\operatorname{Im}\Sigma^R
  &=\frac{-\operatorname{Im}\mathcal T^R
    +\operatorname{Im}\mathcal G_0^R\,|\mathcal T^R|^2}
    {|1+\mathcal G_0^R\mathcal T^R|^2}.
\end{aligned}
\label{eq:sfl_tmatrix_matching}
\end{equation}
At the decoupled fixed point described by $H_*$, the purely elastic amplitude obeys the
optical theorem $\operatorname{Im}\mathcal T^R=\operatorname{Im}\mathcal G_0^R\,|\mathcal T^R|^2$ 
\cite{taylorbook}.
Its finite scattering cross section therefore cancels from the numerator
of the interaction damping, which measures inelastic scattering
\cite{Zarand2004,Mehta2005,Koller2005}.

\subsection{The finite-temperature one-loop function}

\begin{figure}[tb]
  \centering
  \includegraphics{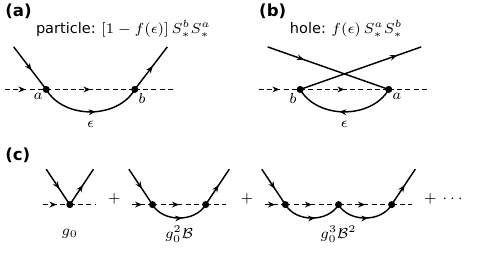}
  \caption{
  Exchange scattering from the residual spin and leading-logarithmic resummation.
  Solid lines represent a screening-channel electron; the dashed line represents
  the free residual spin $\mathbf S_*$, with $S_*=S-1/2$.
  Time runs to the right, and the spin line carries no energy.
  The vertices carry spin components $S_*^a$ and $S_*^b$: vertex $a$ absorbs
  the incoming electron and vertex $b$ emits the outgoing one.
  (a) Particle process: the intermediate state at energy $\epsilon$ is empty,
  with weight $1-f(\epsilon)$ and spin-operator order $S_*^bS_*^a$.
  (b) Hole process: the intermediate state is occupied, with weight $f(\epsilon)$;
  the electron lines cross, and the order is $S_*^aS_*^b$.
  Here $f$ is the Fermi function. Each diagram contains one electron--spin loop.
  The commutator part of their sum defines the one-loop function $\mathcal B$
  of Eq.~\eqref{eq:sfl_loop_integral}, Kondo's logarithm at finite frequency and temperature.
  (c) Dimensionless coefficients of the leading-logarithmic exchange vertex
  in Eq.~\eqref{eq:sfl_loop_series}; the common operator factor
  $\boldsymbol\sigma\cdot\mathbf S_*/(2\nu_*)$ is omitted.
  This panel is schematic: each loop stands for the combination of (a) and (b).
  }
  \label{fig:sfl_loop}
\end{figure}

We now calculate the spin-averaged retarded $T$-matrix
$\mathcal T_{\rm r}^R(\omega,T)$ for scattering of screening-channel
electrons from the residual spin, using the exchange-vertex resummation
shown in Fig.~\ref{fig:sfl_loop}(c). We define it by
\begin{equation}
  G_{\rm low}=\mathcal G_*
    +\mathcal G_*\mathcal T_{\rm r}\mathcal G_*,
  \label{eq:sfl_residual_tmatrix}
\end{equation}
where $G_{\rm low}$ and $\mathcal G_*$ are the local screening-channel
propagators of the field $\psi_*$ for $H_{\rm low}$ and $H_*$,
respectively. Their reference density of states is
$\nu_*(\epsilon)=-\operatorname{Im}\mathcal G_*^R(\epsilon)/\pi$,
with $\nu_*=\nu_*(0)$.
The fixed-point elastic scattering is already included in $\mathcal G_*$.

The logarithms originate from second-order exchange scattering
\cite{Kondo1964,Hewsonbook}, with intermediate electron energy $\epsilon$
and external energy $\omega$ measured from the Fermi level.
Figure~\ref{fig:sfl_loop}(a,b) shows the two time orderings.
The residual spin is free, so a spin flip costs no energy; with the
hole-process fermion sign included, both denominators are $\omega-\epsilon$.

With the vertex ordering of Fig.~\ref{fig:sfl_loop}, the particle and hole
processes carry weights $1-f(\epsilon)$ and $f(\epsilon)$. Their spin factors sum to
\begin{equation}
\begin{aligned}
  &[1-f(\epsilon)]\,S_*^bS_*^a+f(\epsilon)\,S_*^aS_*^b\\
  &\qquad=\frac12\{S_*^a,S_*^b\}
   +\frac12\tanh\frac{\epsilon}{2T}\,[S_*^b,S_*^a].
\end{aligned}
\label{eq:sfl_spin_order}
\end{equation}
Here $1-2f(\epsilon)=\tanh(\epsilon/2T)$.
The anticommutator describes
potential scattering, for which the logarithms from $\epsilon>0$ and
$\epsilon<0$ cancel. The commutator,
$[S_*^b,S_*^a]=i\sum_c\varepsilon_{bac}S_*^c$, has the exchange form and
renormalizes $J_{\rm r}$.
Its weight changes sign together with the
energy denominator, so the two halves of the band add to a logarithm.
The smeared Fermi step cuts this logarithm off at $|\epsilon|\sim T$.

Approximating the metallic screening-channel density of states by its
finite Fermi-level value $\nu_*$ for $|\epsilon|<D_*$
[Eq.~\eqref{eq:sfl_metallic_bath}], we obtain the second-order amplitude
\begin{equation}
\begin{aligned}
  \mathcal V_{\alpha\beta}^{(2)}(z,T)
  &=\frac{J_{\rm r}^2\nu_*}{4}\int_{-D_*}^{D_*}
    \frac{d\epsilon}{z-\epsilon}\sum_{ab}(\sigma^b\sigma^a)_{\alpha\beta}\\
  &\quad\times\bigl\{[1-f(\epsilon)]S_*^bS_*^a
    +f(\epsilon)S_*^aS_*^b\bigr\}.
\end{aligned}
\label{eq:sfl_second_order_vertex}
\end{equation}
The prefactor contains two vertices $J_{\rm r}/2$ and the per-spin state
measure $\nu_*d\epsilon$; the matrix product sums the intermediate
electron spin. Here $\alpha$ labels the outgoing spin and $\beta$ the
incoming one. Using
$\sigma^b\sigma^a=\delta_{ab}\mathbf1+i\sum_c\varepsilon_{bac}\sigma^c$
gives
$\sum_{ab}\sigma^b\sigma^a[S_*^b,S_*^a]=-2\boldsymbol\sigma\cdot\mathbf S_*$.
Thus the commutator part of Eq.~\eqref{eq:sfl_second_order_vertex} is
\begin{equation*}
\begin{aligned}
  \mathcal V_{\rm ex}^{(2)}
  &=\frac{J_{\rm r}^2\nu_*}{4}\boldsymbol\sigma\cdot\mathbf S_*
    \int_{-D_*}^{D_*}d\epsilon\,
    \frac{\tanh(\epsilon/2T)}{\epsilon-z}\\
  &=\mathcal V^{(1)}g_0\mathcal B(z,T),\qquad
    \mathcal V^{(1)}=\frac{J_{\rm r}}2\boldsymbol\sigma\cdot\mathbf S_*.
\end{aligned}
\end{equation*}
Here $g_0=\nu_*J_{\rm r}$. The loop function is
\begin{equation}
  \mathcal B(z,T)=\frac12\int_{-D_*}^{D_*}
  \frac{\tanh(\epsilon/2T)}{\epsilon-z}\,d\epsilon,
  \qquad \operatorname{Im}z>0.
  \label{eq:sfl_loop_integral}
\end{equation}
For the retarded amplitude, $z\to\omega+i0^+$, and we write
$\mathcal B^R(\omega,T)\equiv\mathcal B(\omega+i0^+,T)$.
Iterating the loop, as in Fig.~\ref{fig:sfl_loop}(c), gives the
leading-logarithmic exchange vertex \cite{Abrikosov1965}
\begin{equation}
\begin{aligned}
  \mathcal V_{\alpha\beta}^R(\omega,T)
  &\simeq\frac{\boldsymbol\sigma_{\alpha\beta}\cdot\mathbf S_*}{2\nu_*}
    \bigl[g_0+g_0^2\mathcal B^R\\
  &\qquad\qquad+g_0^3(\mathcal B^R)^2+\cdots\bigr]\\
  &=\frac{\boldsymbol\sigma_{\alpha\beta}\cdot\mathbf S_*}{2\nu_*}
    \frac{g_0}{1-g_0\mathcal B^R}.
\end{aligned}
  \label{eq:sfl_loop_series}
\end{equation}
The spin-averaged $\mathcal T_{\rm r}^R$ starts at second order because
$\langle\mathbf S_*\rangle=0$. The loop $\mathcal B$ replaces the RG time
$\ell$ of Eq.~\eqref{eq:sfl_one_loop_flow}: its real part is the logarithm
cut off at $E\sim\max\{|\omega|,T\}$, and its imaginary part comes from
on-shell intermediate states.
Using $(u-i0^+)^{-1}=\mathcal P/u+i\pi\delta(u)$, where $\mathcal P$
denotes the Cauchy principal value and $\delta$ the Dirac delta, gives,
inside the cutoff,
\begin{equation}
\begin{aligned}
  \operatorname{Im}\mathcal B^R(\omega,T)
    &=q(\omega,T)=\frac\pi2\tanh\frac{\omega}{2T},\\
  \mathcal B^R(0,T)
    &=\int_0^{D_*}\frac{\tanh(\epsilon/2T)}{\epsilon}\,d\epsilon
      \simeq\ln\frac{2e^{\gamma_E}D_*}{\pi T}.
\end{aligned}
\label{eq:sfl_loop_limits}
\end{equation}
Here $\gamma_E$ is Euler's constant.
To evaluate the frequency dependence, define the subtracted loop
$\delta\mathcal B(z,T)=\mathcal B(z,T)-\mathcal B^R(0,T)$.
Parity gives
\begin{equation}
  \delta\mathcal B(z,T)
  =z^2\int_0^{D_*}d\epsilon\,
  \frac{\tanh(\epsilon/2T)}{\epsilon(\epsilon^2-z^2)}.
  \label{eq:sfl_subtracted_loop}
\end{equation}
Extending the upper limit to infinity changes the result by
$O(z^2/D_*^2)$. We use the Matsubara expansion
$\tanh(\epsilon/2T)=4\epsilon T\sum_{n\ge0}(\epsilon^2+\nu_n^2)^{-1}$,
where $\nu_n=(2n+1)\pi T$. Writing $z^2/\epsilon(\epsilon^2-z^2)=\epsilon/(\epsilon^2-z^2)-1/\epsilon$,
and integrating each rational term, produces the convergent sum below,
with $\psi(w)=d\ln\Gamma_{\!E}(w)/dw$ the digamma function and
$\Gamma_{\!E}$ the Euler gamma function:
\begin{equation}
\begin{aligned}
  \delta\mathcal B(z,T)
  &\simeq\sum_{n=0}^\infty
  \left[\frac{1}{n+1/2-iz/(2\pi T)}-\frac{1}{n+1/2}\right]\\
  &=-\psi\!\left(\frac12-\frac{iz}{2\pi T}\right)
    +\psi\!\left(\frac12\right).
\end{aligned}
\label{eq:sfl_loop_sum}
\end{equation}
Since $\psi(1/2)=-\gamma_E-2\ln2$, we obtain, to the stated cutoff accuracy,
\begin{equation}
\begin{aligned}
  \mathcal B^R(\omega,T)&=\ln\frac{D_*}{2\pi T}
    -\psi\!\left(\frac12-\frac{i\omega}{2\pi T}\right)\\
  &=\mathcal L(\omega,T)-\frac{1}{|g_0|}+iq(\omega,T),\\
  \mathcal L(\omega,T)&=\ln\frac{T_{\rm s}}{2\pi T}
    -\operatorname{Re}\psi\!\left(\frac12-\frac{i\omega}{2\pi T}\right),
\end{aligned}
\label{eq:sfl_loop_result}
\end{equation}
with $T_{\rm s}=D_*e^{1/|g_0|}$ and $q$ of Eq.~\eqref{eq:sfl_loop_limits}. This $\mathcal L$
is the thermal logarithm of Eq.~\eqref{eq:sfl_thermal}. Since $g_0<0$, the denominator of
Eq.~\eqref{eq:sfl_loop_series} is $1-g_0\mathcal B^R=|g_0|(\mathcal L+iq)$: $\mathcal L$ is the inverse
running coupling $1/|g|$ of Eq.~\eqref{eq:sfl_one_loop_flow}, cut off at
$\max\{|\omega|,T\}$.

For $S_*=1/2$, we first calculate the Born rate in the constant-DOS
approximation, for $|\omega|,T\ll D_*$. The spin average of the squared
exchange vertex contains the factor
$\langle(\mathbf S_*\cdot\boldsymbol\sigma/2)^2\rangle
=S_*(S_*+1)/4=3/16$. Multiplying by $J_{\rm r}^2$ and the on-shell
electronic factor $\pi\nu_*$ gives
$-\operatorname{Im}\mathcal T_{\rm r}^R=3\pi\nu_*J_{\rm r}^2/16
=3\pi g_0^2/(16\nu_*)$.
A spin flip costs no energy, and the particle and hole processes of
Fig.~\ref{fig:sfl_loop}(a,b) contribute on shell with total weight
$[1-f(\omega)]+f(\omega)=1$. The Born rate therefore depends on
neither $\omega$ nor $T$.

The geometric resummation of Eq.~\eqref{eq:sfl_loop_series} dresses the
exchange vertex by replacing $g_0$ with $g_0/(1-g_0\mathcal B^R)$
\cite{CabreraCano2013}. Squaring the scattering amplitude then replaces
$g_0^2$ in the Born rate with $|g_0/(1-g_0\mathcal B^R)|^2$.
To express this result through the on-shell part of the loop, we use
$\operatorname{Im}[\mathcal B^R/(1-g_0\mathcal B^R)]
=q/|1-g_0\mathcal B^R|^2$.
Here $q=(\pi/2)\tanh(\omega/2T)$ carries the commutator weight
$[1-f(\omega)]-f(\omega)$; multiplying by $\coth(\omega/2T)$ restores
the total particle--hole weight, since $q\coth(\omega/2T)=\pi/2$.
Combining the Born prefactor with the dressed vertex thus gives, within
this resummation,
\begin{equation}
\begin{aligned}
  -\operatorname{Im}\mathcal T_{\rm r}^R
  &\simeq\frac{3\pi}{16\nu_*}
    \left|\frac{g_0}{1-g_0\mathcal B^R}\right|^2\\
  &=\frac{3g_0^2}{8\nu_*}\coth\frac{\omega}{2T}\,
    \operatorname{Im}\frac{\mathcal B^R}{1-g_0\mathcal B^R},
\end{aligned}
  \label{eq:sfl_residual_damping}
\end{equation}
with the $\omega=0$ value understood by continuity.
This is the resummed damping of Ref.~\cite{CabreraCano2013}, with
$\mathcal B$ in place of their electron--Majorana bubble, which they
evaluated at $T=0$.
Using $|g_0/(1-g_0\mathcal B^R)|^2=1/(\mathcal L^2+q^2)$ gives the
frequency and temperature dependence of the residual damping.
Matching the screening-channel amplitude to the physical impurity
$T$-matrix and then to $\Sigma$ through Eq.~\eqref{eq:sfl_tmatrix_matching}
introduces model-dependent factors. We include them in $A_\Sigma>0$, the
leading residual self-energy amplitude, which has dimensions of energy.
Its microscopic matching value is not determined by the low-energy
calculation. Within the resummation above,
\begin{equation}
  \Gamma(\omega,T)
  =\frac{A_\Sigma}{\mathcal L^2+q^2}
  =\frac{A_\Sigma}{\mathcal L^2}
    +O(A_\Sigma\mathcal L^{-4}).
  \label{eq:sfl_rpa_damping}
\end{equation}
At $T=0$ it gives the slow $1/\ln^2$ approach to the Fermi level found
numerically for underscreened impurities \cite{Koller2005}.

The real part of the self-energy, which produces the thermal mass
renormalization of Sec.~\ref{sec:spectral_optical}, follows from
analyticity. With $L_T=\mathcal L(0,T)$, Eq.~\eqref{eq:sfl_loop_result}
gives $\mathcal L+iq=L_T+\delta\mathcal B^R(\omega,T)$, so
Eq.~\eqref{eq:sfl_rpa_damping} reads
$\Gamma=A_\Sigma/|L_T+\delta\mathcal B^R|^2$. At fixed
$x=\omega/(2\pi T)$, $\delta\mathcal B^R$ is independent of $T$ to the
cutoff accuracy of Eq.~\eqref{eq:sfl_loop_sum}. Expanding the denominator
for large $L_T$ gives
\begin{equation*}
  \Gamma=\frac{A_\Sigma}{L_T^2}
    -\frac{2A_\Sigma\operatorname{Re}\delta\mathcal B^R}{L_T^3}
    +O(A_\Sigma L_T^{-4}).
\end{equation*}
Since $\delta\mathcal B$ is analytic for $\operatorname{Im}z>0$,
$i\delta\mathcal B$ supplies the required Kramers--Kronig pair:
$\operatorname{Im}(i\delta\mathcal B^R)=\operatorname{Re}\delta\mathcal B^R$
and $\operatorname{Re}(i\delta\mathcal B^R)=-q$. Thus
\begin{equation}
\begin{aligned}
  \Sigma_{\rm sing}(z,T)
    &\simeq-\frac{iA_\Sigma}{L_T^2}
     +\frac{2iA_\Sigma}{L_T^3}\,\delta\mathcal B(z,T),\\
  \operatorname{Re}\Sigma^R_{\rm sing}(\omega,T)
    &\simeq-\frac{2A_\Sigma q}{L_T^3}\\
    &=-\frac{\pi A_\Sigma\tanh(\omega/2T)}{L_T^3}.
\end{aligned}
\label{eq:sfl_real_part}
\end{equation}
This is a low-energy expansion at fixed $\omega/T$. The subscript ``sing'' denotes its singular
part; an additive real constant and regular analytic terms depend on
the self-energy outside this low-energy window.

\subsection{Metallic self-consistency at finite filling}

DMFT self-consistency preserves the impurity logarithms on the metallic
branch, as also found in orbital-selective Mott phases
\cite{Greger2013,Kugler2019}.

Let $\Sigma_0=\operatorname{Re}\Sigma^R(0,0)$ and
$\epsilon_F^*=\mu(0)-\Sigma_0$, and assume $|\epsilon_F^*|<D$.
Writing $\zeta=\omega+\mu(T)-\Sigma^R(\omega,T)$, define the Hilbert
transform $\mathcal H(\zeta)=\int_{-D}^{D}d\epsilon\,
\nu_0(\epsilon)/(\zeta-\epsilon)$. For the semicircular DOS,
\begin{equation}
\begin{aligned}
  \mathcal H(\zeta)&=\frac{2}{D^2}
    \left[\zeta-\sqrt{\zeta^2-D^2}\right],\\
  G^R&=\mathcal H(\epsilon_F^*)
    +\mathcal H'(\epsilon_F^*)\,\delta\zeta+O(\delta\zeta^2),\\
  \delta\zeta&=\omega+\mu(T)-\mu(0)\\
    &\quad-[\Sigma^R(\omega,T)-\Sigma_0].
\end{aligned}
\label{eq:sfl_bethe_expansion}
\end{equation}
The square root is on the retarded branch, chosen so that
$\mathcal H(\zeta)\sim1/\zeta$ at large $|\zeta|$ and
$\operatorname{Im}\mathcal H<0$ for $\operatorname{Im}\zeta>0$.
The prime denotes differentiation with respect to $\zeta$; values at
$\epsilon_F^*$ mean limits from the upper half-plane.
As $T\to0$, $\delta\zeta(0,T)\to0$ by the definition of $\Sigma_0$ and since
$\Gamma(0,T)=A_\Sigma/L_T^2\to0$. Thus
$G^R(0,0)=\mathcal H(\epsilon_F^*+i0^+)$, whose imaginary part is
$-\pi\nu_0(\epsilon_F^*)$. With $\Delta=(D^2/4)G$
[Eq.~\eqref{eq:dmft_bethe}], the bath width is finite:
\begin{equation}
  -\operatorname{Im}\Delta^R(0,0)
    =\frac12\sqrt{D^2-(\epsilon_F^*)^2}>0.
  \label{eq:sfl_metallic_bath}
\end{equation}
The correction induced by residual scattering is
$\delta\Sigma\equiv\Sigma^R-\Sigma_0=O(g^2)$, with a fixed energy
prefactor understood. Away from the band edges, $\mathcal H'$ is finite.
Writing $\delta G=G^R-G^R(0,0)$ and
$\delta\Delta=\Delta^R-\Delta^R(0,0)$, the expansion in
Eq.~\eqref{eq:sfl_bethe_expansion} gives
$\delta G=O(g^2)$ and $\delta\Delta=(D^2/4)\delta G=O(g^2)$.
This assumes
a smooth response at fixed filling, including the adjustment of $\mu$;
regular corrections proportional to the small external energy or
temperature do not change the logarithmic counting. The expansion is
in $\delta\zeta$ and does not require a finite $\partial_\omega\Sigma$.

The Weiss propagator and the effective screening-channel density of
states are smooth functions of this finite bath. Consequently the
relative change $\delta\nu_*/\nu_*$ is $O(g^2)$. The limiting
$\nu_*$ normalizes $g$; it need not equal the bare-band DOS
$\nu_0(\epsilon_F^*)$. In the one-loop term the bath correction thus
appears as $g^2[1+O(g^2)]$, changing $dg/d\ell$ only at order $g^4$.
It changes neither the leading inverse logarithm nor the cubic
beta-function coefficient used below.

\subsection{From the transport bubble to an explicit resistivity}

At fixed frequency, the spectral function in Eq.~\eqref{eq:transport_spectrum} is a Lorentzian
in the bare band energy $\epsilon$, with half-width $\Gamma(\omega,T)$ and center
$\xi(\omega,T)=\omega+\mu(T)-\operatorname{Re}\Sigma^R(\omega,T)$.
For a center away from the band edges and a smooth $\Phi$, the
band-energy integral is controlled by this narrow Lorentzian
\cite{Berthod2013}. Using
$\int_{-\infty}^{\infty}du\,\Gamma^2/
[\pi^2(u^2+\Gamma^2)^2]=1/(2\pi\Gamma)$ and defining
$\Phi_*=\Phi(\epsilon_F^*)$ gives
\begin{equation}
\begin{aligned}
  \mathcal K(\omega,T)
  &=\frac{\Phi(\xi)}{2\pi\Gamma}
    +\frac{\Gamma}{4\pi}\Phi''(\xi)+\cdots\\
  &\simeq\frac{\Phi_*}{2\pi\Gamma}.
\end{aligned}
\label{eq:sfl_band_integral}
\end{equation}
As $T\to0$, both $\Gamma$ and $\xi-\epsilon_F^*$ vanish within the thermal
window, so the second line of Eq.~\eqref{eq:sfl_band_integral} becomes
asymptotically exact.

Define the normalized thermal average of a function $X(\omega,T)$ by
$\langle X\rangle_T=\int_{-\infty}^{\infty}d\omega\,
[-f'(\omega)]X(\omega,T)$, where the prime denotes a frequency derivative.
The transport relations \eqref{eq:transport_moments}--\eqref{eq:dc_transport} become
\begin{equation}
  \frac{\rho}{\rho_0}
    \simeq\frac{2}{\pi\Phi_*\langle\Gamma^{-1}\rangle_T}
     =\frac{A_\rho}{\langle\mathcal L^2\rangle_T},\\
\label{eq:sfl_harmonic_average}
\end{equation}
with the dimensionless amplitude $A_\rho=\frac{2A_\Sigma}{\pi\Phi_*}$. We used $\Gamma=A_\Sigma/\mathcal L^2$ as the working
one-loop ansatz. 

The thermal average can be performed analytically. Set
\begin{equation}
\begin{alignedat}{2}
  x&=\frac{\omega}{2\pi T}, &\qquad
  p(x)&=\frac\pi2\operatorname{sech}^2(\pi x),\\
  h(x)&=\operatorname{Re}\psi(1/2+ix), &
  \lambda_T&=\ln\frac{T_{\rm s}}{2\pi T}.
\end{alignedat}
\label{eq:sfl_thermal_measure}
\end{equation}
Then $[-f'(\omega)]d\omega=p(x)dx$, $\int p(x)dx=1$, and
$\mathcal L=\lambda_T-h(x)$. Angle brackets below denote averaging
with $p(x)$. To obtain the mean and variance of $h$,
use Barnes's beta integral, the real-line form of Barnes's first lemma
\cite{Barnes1908,DLMFBarnes},
\begin{equation}
\begin{aligned}
  &\mathcal I(a,b,c,d)
  \equiv\int_{-\infty}^{\infty}\frac{dx}{2\pi}
    \Gamma_{\!E}(a+ix)\Gamma_{\!E}(b+ix)\\
  &\quad\times\Gamma_{\!E}(c-ix)\Gamma_{\!E}(d-ix)\\
  &\quad=\frac{\Gamma_{\!E}(a+c)\Gamma_{\!E}(a+d)
    \Gamma_{\!E}(b+c)\Gamma_{\!E}(b+d)}
    {\Gamma_{\!E}(a+b+c+d)}.
\end{aligned}
\label{eq:sfl_barnes}
\end{equation}
All four parameters have positive real parts. At $a=b=c=d=1/2$,
$\mathcal I=1$ and $|\Gamma_{\!E}(1/2+ix)|^4/(2\pi)=p(x)$.
As for Euler's beta integral, differentiating a gamma factor inserts
its digamma function. Thus
$\partial_a\mathcal I=2\psi(1)-\psi(2)=-\gamma_E-1\equiv m$.
Using $\psi'(1)=\pi^2/6$ and $\psi'(2)=\pi^2/6-1$, mixed derivatives give
$\partial_a\partial_b\mathcal I
=\langle\psi(1/2+ix)^2\rangle=m^2+1-\pi^2/6$ and
$\partial_a\partial_c\mathcal I
=\langle|\psi(1/2+ix)|^2\rangle=m^2+1$.
Taking half their sum yields $\langle h^2\rangle$; the imaginary parts
average to zero by parity. Consequently,
\begin{equation}
\begin{aligned}
  \langle h\rangle&=-\gamma_E-1,\qquad
  \operatorname{Var}(h)=1-\frac{\pi^2}{12},\\
  \langle\mathcal L^2\rangle_T
    &=(\lambda_T+\gamma_E+1)^2+1-\frac{\pi^2}{12}.
\end{aligned}
\label{eq:sfl_digamma_moments}
\end{equation}
Inserting this result into Eq.~\eqref{eq:sfl_harmonic_average} gives
\begin{equation}
\begin{aligned}
  \frac{\rho_{\text{1-loop}}(T)}{\rho_0}
  &\simeq\frac{A_\rho}
    {\ln^2(T_\rho^{(1)}/T)+1-\pi^2/12},\\
 \end{aligned}
  \label{eq:sfl_one_loop_rho}
\end{equation}
with  $T_\rho^{(1)}=\frac{e^{\gamma_E+1}}{2\pi}T_{\rm s}$. At low temperatures
the nonuniversal part $1-\pi^2/12$ may be omitted.

\subsection{Two-loop running and the fitting formula}
\label{app:two_loop}

For a single screening channel, the beta function is
$dg/d\ln E=-g^2+g^3/2+O(g^4)$
\cite{AbrikosovMigdal1970,Fowler1971,Nozieres1980,Fritz2004}.
For the ferromagnetic residual exchange, use the positive inverse
coupling $Y=-1/g$. With $\ell=\ln(D_*/E)$ increasing as the energy
is lowered, the flow becomes
\begin{equation}
  \frac{dY}{d\ell}=1+\frac{1}{2Y}+O(Y^{-2}).
  \label{eq:sfl_two_loop_flow}
\end{equation}
Keeping the first two terms and integrating gives
\begin{equation}
  Y-\frac12\ln\!\bigl(Y+\tfrac12\bigr)=\ell+C_0.
\label{eq:sfl_integrated_flow}
\end{equation}
The value of $Y$ at the cutoff $D_*$ fixes $C_0$. At large $\ell$,
\begin{equation}
  Y=\ell+\frac12\ln\ell+C_0+O\!\left(\frac{\ln\ell}{\ell}\right).
\label{eq:sfl_loglog_expansion}
\end{equation}

To relate $Y$ to the resistivity, evaluate the running exchange at $E=T$.
At fixed $x=\omega/(2\pi T)$, the damping is
$\Gamma=A_\Sigma Y^{-2}[1+F_1(x)/Y+O(Y^{-2})]$, where $F_1$ is the
finite frequency-dependent correction after the large logarithms have
been absorbed into $Y$. Define the inverse coupling inferred from
resistivity as $y=\sqrt{A_\rho\rho_0/\rho}$. The conductivity measures
the thermal average of the inverse damping, so
Eq.~\eqref{eq:sfl_harmonic_average} gives
$y^2=A_\Sigma\langle\Gamma^{-1}\rangle_T$ within the transport
approximation. Expanding the inverse damping and taking the positive
square root gives
\begin{equation}
\begin{aligned}
  y^2 &= Y^2-Y\langle F_1\rangle+O(1),\\
  y &= Y-\frac12\langle F_1\rangle+O(Y^{-1}).
\end{aligned}
\label{eq:sfl_observable_matching}
\end{equation}
Here $\langle F_1\rangle=\int dx\,p(x)F_1(x)$ uses the normalized
weight of Eq.~\eqref{eq:sfl_thermal_measure} and is independent of $T$.
Thus thermal averaging shifts the inverse coupling by a constant to
leading order. In Eq.~\eqref{eq:sfl_integrated_flow}, this shift changes
the linear term by $-\langle F_1\rangle/2$ and the logarithm by
$O(Y^{-1})$. At $E=T$ this gives
\begin{equation*}
  y-\frac12\ln\!\bigl(y+\tfrac12\bigr)
  =\ln\frac{D_*}{T}+C_0-\frac12\langle F_1\rangle+O(Y^{-1}).
\end{equation*}
Adding a constant to $\ln(D_*/T)$ amounts to rescaling $D_*$.
Absorbing the constants into $T_\rho$ and dropping the $O(Y^{-1})$
remainder gives the two-loop fitting formula \eqref{eq:sfl_rho_fit}.
With $\ell_\rho=\ln(T_\rho/T)$, the same expansion as in
Eq.~\eqref{eq:sfl_loglog_expansion} gives
$y=\ell_\rho+\tfrac12\ln\ell_\rho+O(\ln\ell_\rho/\ell_\rho)$.
Expanding $A_\rho/y^2$ then yields
\begin{equation}
  \frac{\rho(T)}{\rho_0}
  =\frac{A_\rho}{\ell_\rho^2}
  \left[1-\frac{\ln\ell_\rho}{\ell_\rho}
  +O\!\left(\frac{\ln^2\ell_\rho}{\ell_\rho^2}\right)\right].
  \label{eq:sfl_rho_asymptote}
\end{equation}

\subsection{Larger spins}
\label{app:larger_spins}

For $S\geq3/2$, a single channel still screens only half a unit of spin
\cite{Nozieres1980,Coleman2003,Mehta2005}.
Equation~\eqref{eq:sfl_effective_hamiltonian} then holds with
$S_*=S-1/2\geq1$.
Most of the derivation carries over. The fixed-point scattering
remains elastic, so the optical-theorem cancellation below
Eq.~\eqref{eq:sfl_tmatrix_matching} is unchanged. The one-loop
algebra of Eq.~\eqref{eq:sfl_spin_order} uses only the commutation
relations of $\mathbf S_*$, so $\mathcal B$, the thermal logarithm
$\mathcal L$, and the unit one-loop coefficient do not depend on $S_*$.
The two-loop coefficient $1/2$ in Eq.~\eqref{eq:sfl_two_loop_flow}
depends only on the number of screening channels \cite{Nozieres1980}.

Spin averaging the Born rate gives
$\langle(\mathbf S_*\cdot\boldsymbol\sigma)^2\rangle=S_*(S_*+1)$, so
$A_\Sigma$ and $A_\rho$ acquire this factor together with
model-dependent matching factors. The resummation derived for $S_*=1/2$
in Ref.~\cite{CabreraCano2013} extends to arbitrary $S_*$ at
leading-logarithmic accuracy through Eq.~\eqref{eq:sfl_loop_series}.
The resistivity formulas \eqref{eq:sfl_one_loop_rho} and
\eqref{eq:sfl_rho_fit} retain their form, with spin-dependent $A_\rho$ and $T_\rho$.

At fixed residual exchange field, increasing $S_*$ slows the logarithmic
flow. The logarithms come from the commutator $[S_*^b,S_*^a]$, which
is linear in $S_*$, whereas the strength of the residual-spin scattering
is set by the dimensionless exchange field $c=|g(T)|S_*$.
Equation~\eqref{eq:sfl_two_loop_flow} gives
\begin{equation}
  \frac{d\ln c}{d\ln T}=\frac{c}{S_*}+\frac{c^2}{2S_*^2}+\cdots.
  \label{eq:sfl_classical_flow}
\end{equation}
At fixed $c$, the leading logarithmic flow rate scales as $1/S_*$,
as does $d\ln\rho/d\ln T$. In the Born regime,
$d\ln\rho/d\ln T\simeq2|g(T)|=2c/S_*$.

\section{Coherent-potential description of ultrahigh-temperature transport}
\label{app:highT_cpa}

We specialize the paramagnetic many-body coherent-potential approximation
(CPA) of Ref.~\cite{Edwards1999}, which extends Kubo's CPA for random
spin systems \cite{Kubo1974},
to $T\gg D,J(S+1/2)$ \cite{Rangette1973} at fixed filling $0<n<2$.
The atomic Green function is exact in this
limit; its CPA embedding is an approximation to the finite-spin lattice
problem. Throughout this appendix, $z=E+i0^+$ is measured from the bare
band center, with $E=\omega+\mu$.

\subsection{Atomic transitions and lattice embedding}

The empty and doubly occupied atomic states have zero exchange energy.
For single occupancy, coupling the electron spin to $S$ gives total spins
$S+1/2$ and $S-1/2$, with energies and relative multiplicities
\begin{equation}
\begin{alignedat}{2}
 E_+&=\frac{JS}{2}, &\qquad w_+&=\frac{S+1}{2S+1},\\
 E_-&=-\frac{J(S+1)}{2}, & w_-&=\frac{S}{2S+1}.
\end{alignedat}
\label{eq:cpa_atomic_multiplets}
\end{equation}
At infinite temperature, each conduction spin-orbital is occupied
independently with probability $p=n/2$, and the $2S+1$ local-spin states
have equal weight. Each atomic state with $N=0,1,2$ electrons therefore has
probability $P_N=p^N(1-p)^{2-N}/(2S+1)$. For spin $\uparrow$, the
Lehmann representation is
\begin{equation}
 G_S^{\rm at}(z)=\sum_{\substack{m,n\\N_n=N_m+1}}
 \frac{(P_{N_m}+P_{N_n})
 |\langle n|c_\uparrow^\dagger|m\rangle|^2}
 {z-(\mathcal E_n-\mathcal E_m)}.
 \label{eq:cpa_atomic_lehmann}
\end{equation}
Here $m,n$ label atomic eigenstates, $N_m$ is the charge of state $m$,
and $\mathcal E_m$ is its exchange energy, excluding $-\mu N_m$ because
$z$ is measured from the band center. The probability sums are
$P_0+P_1=(1-p)/(2S+1)$ and $P_1+P_2=p/(2S+1)$.

To evaluate the matrix elements, let $M=-S,-S+1,\ldots,S$ be the local
spin projection in the empty or doubly occupied state $|N,M\rangle$,
with $N=0,2$. The single-occupancy
states $|\lambda,m_F\rangle$ have total spin $S+1/2$ for $\lambda=+$
and $S-1/2$ for $\lambda=-$, with total projection $m_F$.
Spin addition gives the following squared matrix elements for adding
a spin-up electron:
\begin{equation*}
\begin{array}{c@{\quad}cc}
 &\lambda=+&\lambda=-\\
 0\to1 &\dfrac{S+M+1}{2S+1}&\dfrac{S-M}{2S+1}\\[5pt]
 1\to2 &\dfrac{S-M+1}{2S+1}&\dfrac{S+M}{2S+1}
\end{array}
\end{equation*}
The first row is
$|\langle\lambda,M+1/2|c_\uparrow^\dagger|0,M\rangle|^2$;
the second is
$|\langle2,M|c_\uparrow^\dagger|\lambda,M-1/2\rangle|^2$.
Summing over $M$, using $\sum_M M=0$, gives $S+1$ and $S$ in the
two columns for either row. Multiplication by the probability sums
therefore gives residues $(1-p)w_\lambda$ for $0\to1$, at energy
$E_\lambda$, and $pw_\lambda$ for $1\to2$, at energy $-E_\lambda$.
Thus the per-spin atomic Green function is
\begin{equation}
 G_S^{\rm at}(z)=\sum_{\lambda=\pm}w_\lambda
 \left[\frac{1-p}{z-E_\lambda}+\frac{p}{z+E_\lambda}\right].
 \label{eq:cpa_atomic_green}
\end{equation}
There are four transitions, at $E_\pm$ and $-E_\pm$.
This is the infinite-temperature case of the atomic Green function of
Refs.~\cite{NoltingMatlak1984,Edwards1999}, which holds for general thermal weights.

The CPA treats these transitions as static random site energies in a
common effective medium. Label the four alloy components by
$\ell=1,\ldots,4$, with energies and probabilities
\begin{equation*}
\begin{aligned}
 (e_1,e_2,e_3,e_4)&=(E_+,E_-,-E_+,-E_-),\\
 (W_1,W_2,W_3,W_4)&=((1-p)w_+,(1-p)w_-,\\
 &\qquad pw_+,pw_-).
\end{aligned}
\end{equation*}
Relative to a medium with local Green function $G$ and self-energy
$\Sigma$, component $\ell$ changes the local potential by $e_\ell-\Sigma$.
Repeated scattering from this difference is a geometric series, giving
its $T$-matrix $\tau_\ell$ and local Green function $G_\ell$:
\begin{equation}
\begin{aligned}
 \tau_\ell&=\frac{e_\ell-\Sigma}{1-(e_\ell-\Sigma)G},\\
 G_\ell&=G+G\tau_\ell G
        =\frac{1}{G^{-1}+\Sigma-e_\ell}.
\end{aligned}
\label{eq:cpa_single_site_scattering}
\end{equation}
The CPA condition is zero average extra scattering,
$\sum_\ell W_\ell\tau_\ell=0$ \cite{Soven1967,Velicky1968}. Since $\sum_\ell W_\ell=1$, this is
equivalent to $\sum_\ell W_\ell G_\ell=G$. With
$\mathcal G_0^{-1}=G^{-1}+\Sigma$, the condition becomes
$G=\sum_\ell W_\ell/(\mathcal G_0^{-1}-e_\ell)
=G_S^{\rm at}(\mathcal G_0^{-1})$.
For the semicircular band, define $t_*^2=D^2/4$ and use the hybridization
$\Delta=t_*^2G$ to close the equations:
\begin{equation}
\begin{aligned}
 \mathcal G_0^{-1}(z)&=z-t_*^2G(z),\\
 G(z)&=G_S^{\rm at}\bigl(\mathcal G_0^{-1}(z)\bigr),\\
 \Sigma(z)&=\mathcal G_0^{-1}(z)-G(z)^{-1}.
\end{aligned}
\label{eq:cpa_closure}
\end{equation}
These are the paramagnetic CPA equations of Ref.~\cite{Edwards1999},
with zero local spin--electron correlation at infinite temperature.
The static-alloy replacement retains the multiplet weights but omits quantum
dynamics, so it is not exact finite-$S$ DMFT. As $S\to\infty$ at fixed $JS$, the four
energies merge into $\pm JS/2$ with equal weights \cite{Edwards1999}, and
Eq.~\eqref{eq:cpa_closure} becomes the DMFT of the classical-spin model
\cite{Furukawa1994}.

\subsection{Transport kernel and linear coefficient}

As in the alloy CPA, vertex corrections to the conductivity vanish for a
single band with local scatterers \cite{Velicky1969}.
For the semicircular DOS and any causal local self-energy, the band-energy
integral in Eq.~\eqref{eq:transport_moments} can be done in closed form.
Set $Z=E+i\eta-\Sigma(E+i\eta)$, $b=\operatorname{Im}Z>0$, and take
the retarded limit $\eta\to0^+$.
The semicircular identity $Z=G(Z)^{-1}+t_*^2G(Z)$ lets us evaluate the
resolvent weighted by $\Phi(\epsilon)=(D^2-\epsilon^2)\nu_0(\epsilon)$:
\begin{equation}
\begin{aligned}
 \mathcal F(Z)&\equiv\int_{-D}^{D}d\epsilon\,
   \frac{\Phi(\epsilon)}{Z-\epsilon}\\
  &=(D^2-Z^2)G+Z=3t_*^2G-t_*^4G^3,\\
 \mathcal F'(Z)&=-3t_*^2G^2.
\end{aligned}
\label{eq:cpa_transport_resolvent}
\end{equation}
The second line follows by polynomial division, using
$\int\nu_0(\epsilon)d\epsilon=1$ and
$\int\epsilon\nu_0(\epsilon)d\epsilon=0$. For the derivative, use
$dG/dZ=-G^2/(1-t_*^2G^2)$, obtained by differentiating the semicircular
identity. The spectral function is
$A(E-\mu,\epsilon)=b/\{\pi[(\operatorname{Re}Z-\epsilon)^2+b^2]\}$.
The integrands of $-\operatorname{Im}\mathcal F/b$ and
$\operatorname{Re}\mathcal F'$ add to
$2\Phi(\epsilon)b^2/[(\operatorname{Re}Z-\epsilon)^2+b^2]^2$.
Therefore the integral of $\Phi A^2$ is
\begin{equation*}
 \mathcal K(E)=\frac{1}{2\pi^2}
 \left[-\frac{\operatorname{Im}\mathcal F(Z)}{b}
       +\operatorname{Re}\mathcal F'(Z)\right].
\end{equation*}
Finally, taking the imaginary part of $Z=G^{-1}+t_*^2G$ gives
$b=-\operatorname{Im}G\,(|G|^{-2}-t_*^2)$. Substitution yields
\begin{equation}
 \mathcal K(E)=\frac{t_*^2(\mathrm{Im}\,G)^2}{\pi^2}
   \frac{3-t_*^2|G|^2}{1-t_*^2|G|^2},
\label{eq:cpa_transport_kernel}
\end{equation}
understood as a retarded limit. Write $\mathcal K_S^\infty(E)$ for the
$T\to\infty$ kernel at fixed $S,J,D,n$, and let
$I_S=\int_{-\infty}^{\infty}dE\,\mathcal K_S^\infty(E)$.
The kernel is dimensionless and $I_S$ has dimensions of energy.
Combining Eqs.~\eqref{eq:dc_transport} and
\eqref{eq:highT_fermi_window} gives the slope below. Its weak-exchange limit,
$J\sqrt{S(S+1)}\ll D$, follows from the Born damping
$\Gamma(E)\simeq\pi\nu_0(E)J^2S(S+1)/4$ and
$\mathcal K(E)\simeq\Phi(E)/[2\pi\Gamma(E)]$:
\begin{gather}
\begin{aligned}
 \frac{\rho}{\rho_0}&\sim a_S^\infty\frac{T}{D},\\
 a_S^\infty&=\frac{4D}{\pi^2n(2-n)I_S},
\end{aligned}
\label{eq:cpa_highT_slope}\\
 a_S^{\rm Born}=\frac{3(J/D)^2S(S+1)}{2n(2-n)}.
\label{eq:cpa_born_slope}
\end{gather}
Inserting the Green function of Eq.~\eqref{eq:cpa_closure} gives the
dimensionless, parameter-free CPA coefficient $a_S^{\rm CPA}$. At larger
exchange, spectral-weight redistribution into the exchange bands suppresses
the Born growth; the coefficient must be computed numerically.

\end{document}